\documentclass[noeprint,noshowpacs,nopreprintnumbers,twocolumn,prb,showpacs,superscriptaddress,amsmath,amssymb,citeautoscript,10pt,aps]{revtex4-1}
\usepackage{graphicx}
\usepackage{dcolumn}
\usepackage{color}
\usepackage{bm}
\usepackage{ulem}
\usepackage[hidelinks]{hyperref}
\hypersetup{
    colorlinks,
    citecolor=blue,
    filecolor=blue,
    linkcolor=blue,
    urlcolor=blue
}

\newcommand{\ud}{\,\mathrm{d}}
\newcommand{\im}{\mathrm{i}}
\newcommand{\e}{\textrm{e}}

\DeclareMathOperator{\C}{\mathcal{C}}
\DeclareMathOperator{\PH}{\mathcal{P}}
\DeclareMathOperator{\TR}{\mathcal{T}}

\DeclareMathOperator{\hside}{\Theta}

\DeclareMathOperator{\Order}{O}

\DeclareMathOperator{\Det}{Det}
\DeclareMathOperator{\Tr}{Tr}

\allowdisplaybreaks

\begin{document}

\title{Quasiperiodic dynamical quantum phase transitions in multiband topological insulators and connections with entanglement entropy and fidelity susceptibility}
\author{T.~Mas{\l}owski}
\affiliation{The Faculty of Mathematics and Applied Physics, Rzesz\'ow University of Technology, al.~Powsta\'nc\'ow Warszawy 6, 35-959 Rzesz\'ow, Poland}
\date{\today}
\author{N.~Sedlmayr}
\email{sedlmayr@umcs.pl}
\affiliation{Institute of Physics, M.~Curie-Sk{\l}odowska University, 20-031 Lublin, Poland}

\begin{abstract}
We investigate the Loschmidt amplitude and dynamical quantum phase transitions in multiband one dimensional topological insulators. For this purpose we introduce a new solvable multiband model based on the Su-Schrieffer-Heeger model, generalized to unit cells containing many atoms but with the same symmetry properties. Such models have a richer structure of dynamical quantum phase transitions than the simple two-band topological insulator models typically considered previously, with both quasiperiodic and aperiodic dynamical quantum phase transitions present. Moreover the aperiodic transitions can still occur for quenches within a single topological phase. We also investigate the boundary contributions from the presence of the topologically protected edge states of this model. Plateaus in the boundary return rate are related to the topology of the time evolving Hamiltonian, and hence to a dynamical bulk-boundary correspondence. We go on to consider the dynamics of the entanglement entropy generated after a quench, and its potential relation to the critical times of the dynamical quantum phase transitions. Finally, we investigate the fidelity susceptibility as an indicator of the topological phase transitions, and find a simple scaling law as a function of the number of bands of our multiband model which is found to be the same for both bulk and boundary fidelity susceptibilities.
\end{abstract}

\maketitle

%\tableofcontents

\section{Introduction}

There are many severe obstacles to formulating a general theory of non-equilibrium processes. Indeed, contrary to equilibrium phenomena, it seems unlikely that this is possible. Nevertheless progress has been made along specific directions by focusing on particular forms of non-equilibrium dynamics, for example, by considering quantum quenches. In a quantum quench the system is prepared in a state, typically an eigenstate of some Hamiltonian, and then time evolved by a Hamiltonian of which it is not an eigenstate. Thus this corresponds to suddenly quenching a parameter of a Hamiltonian. These have not only played a major role in the modern investigation of thermalization of quantum systems,\cite{Rigol2008a,Calabrese2011,Sirker2014a} but also in quantum dynamics\cite{Bloch2008,Polkovnikov2011,Sedlmayr2013}.

A dynamical quantum phase transition is defined in terms of the Loschmidt amplitude, the overlap between an initial state and its time evolved counterpart, $L(t)=\langle\psi_1|\e^{-\im \hbar H_2t}|\psi_1\rangle$, where $|\psi_1\rangle$ is an initial state and $H_2$ is a Hamiltonian.  It occurs when non-analyticities appear in the rate function, $-(\ln|L(t)|)/N$, as a function of time $t$ in the thermodynamic limit $N\to\infty$. They are analogous to the non-analyticities which appear in derivatives of the free energy as a function of a controlling parameter for a standard phase transition. This definition of a dynamical quantum phase transition was first proposed for the Ising chain\cite{Heyl2013,Heyl2018a,Sedlmayr2019a} and soon shown to apply more generally\cite{Karrasch2013,Andraschko2014,Halimeh2018,Mishra2018,Shpielberg2018,Srivastav2019,Zunkovic2018,Hagymasi2019,Huang2019,Jafari2019,Soriente2019,Cao2019,Gurarie2019,Abdi2019}, including in topological insulators.\cite{Vajna2015,Schmitt2015,Jafari2016,Jafari2017a,Sedlmayr2018,Jafari2018,Zache2019} Although it initially appeared that there was a close connection between equilibrium and dynamical quantum phase transitions,\cite{Heyl2013,Karrasch2013,Heyl2014} this has proved not to hold generally.\cite{Vajna2014,Andraschko2014,Vajna2015,Karrasch2017,Jafari2017,Jafari2017a,Jafari2019} Dynamical quantum phase transitions are not confined however to sudden quenches, but can also be seen for example in Floquet systems\cite{Sharma2014,Yang2019} or using slow quenches.\cite{Puskarov2016,Sharma2016,Bhattacharya2017}

\begin{figure}
\includegraphics*[width=0.99\linewidth]{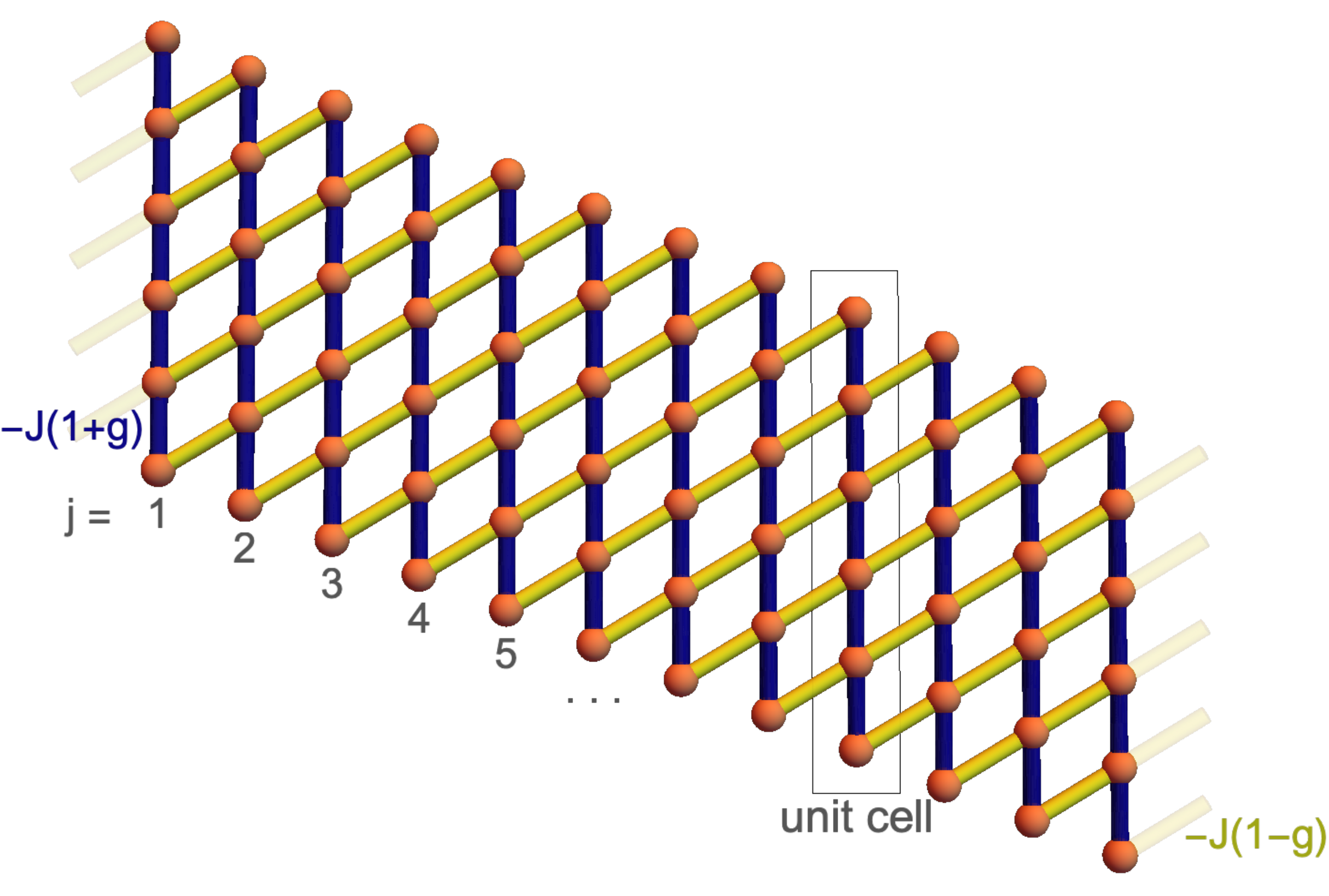}
\caption{A schematic of the generalized SSH model chain. Purple lines have hopping $-J(1+g)$ and yellow lines $-J(1-g)$. For $N=2$ this is equivalent to the SSH model. This model may be further modified by removing some yellow links between unit cells.}
\label{figure_01}
\end{figure}

The quench scenario most typically investigated for dynamical quantum phase transitions starts from a ground state as the initial state, however many generalisations to mixed states, thermal states, and open systems have also been made.\cite{Mera2017,Sedlmayr2018b,Bhattacharya2017a,Heyl2017,Abeling2016,Lang2018,Lang2018a,Kyaw2018} Experimental evidence for dynamical quantum phase transitions has been obtained in ion traps\cite{Jurcevic2017} and old-atom quantum simulators\cite{Smale2019}. A scheme for experimentally observing dynamical quantum phase transitions in a topological nano-mechanical lattice has also been proposed.\cite{Tian2019} The model we introduce below is relatively easy to achieve in such systems as no gauge fields are necessary.

Topological insulators and superconductors are gapped systems where phases are labelled by topological indices defined for bands.\cite{Hasan2010} One of the most celebrated results about topological insulators and superconductors is the existence of their topologically protected edge modes. According to the bulk-boundary correspondence topologically non-trivial phases will have protected states at the edges of the system, typically pinned to zero energy in 1D.\cite{Ryu2002,Teo2010} The number of modes is related to the topological invariant which classifies the topological phase.\cite{Ryu2010} In the context of dynamical quantum phase transitions a bulk-boundary correspondence has also been proposed focusing on boundary contributions to the dynamics.\cite{Sedlmayr2018,Sedlmayr2019a} Elsewhere it has been observed that boundary conditions can potentially be a relevant perturbation for the dynamics of some specific models.\cite{Khatun2019}

Much of the work on the preceding topological insulators has considered two-band models. For multiband topological systems a connection between nodes in the wavefunction overlap and the existence of robust dynamical quantum phase transitions has been demonstrated.\cite{Huang2016,Jafari2019} Further work has been done on multiband models with a singly occupied band.\cite{Mendl2019} Here we introduce a simple, and in the bulk analytically solvable, model for a topological insulator with an arbitrary number of bands. The model nonetheless has a simple topological phase diagram, and the quenches satisfy the criteria of Ref.~\onlinecite{Huang2016} for robustness. This allows us to further investigate the dynamical quantum phase transitions in multiband models, the role of the protected edge states, and their general relation, if any, to entanglement dynamics. We find that this model can have both quasiperiodic and aperiodic dynamical quantum phase transitions.

We consider the return rate of the Loschmidt amplitude in which the dynamical quantum phase transitions appear as cusps at critical times. We find a condition for when both the quasiperiodic and aperiodic critical times appear. Furthermore we demonstrate that the aperiodic, but not quasiperiodic, critical times can also appear for a quench within a single topological phase. Boundary contributions to the return rate show up as plateaus caused by zeroes which appear in the eigenvalues of the Loschmidt matrix.\cite{Sedlmayr2018} These zero eigenvalues between critical times but only for quenches where the time evolving Hamiltonian is topologically non-trivial, and therefore possesses topologically protected edge states. This dynamical bulk-boundary correspondence remains true even for quenches within a single topological phase.

Quench dynamics is not solely concerned with the Loschmidt amplitude. One other quantity which is of interest for topological systems is the dynamics of the entanglement entropy.\cite{Rahmani2010,Sedlmayr2018,Halder2018,Pastori2019} It has been demonstrated, in the context of the Su-Schrieffer-Heeger model,\cite{Su1980} that there is a connection between the zeros of the Loschmidt amplitude, and the oscillations in the entanglement entropy.\cite{Sedlmayr2018} We investigate here if this holds for the more general multiband model considered. Similarly the fidelity, which is the overlap between two states and is therefore clearly related to the Loschmidt amplitude, can be used to analyse topological phase transitions.\cite{Zanardi2007b,Abasto2008,Jozsa1994,You2007,Yang2007,Hamma2008,Sirker2010,Sirker2014,Wang2013a} We find an unexpected relation between the bulk and boundary fidelity susceptibility as a function of the number of bands. For our model all boundary, and all bulk, fidelity susceptibility curves can each be collapsed onto a single curve for the boundary or bulk susceptibility. Furthermore this scaling is unexpectedly the same for both bulk and boundary terms.

This paper is ordered as follows: in section \ref{sec_model} we introduce our generalized Su-Schrieffer-Heeger\cite{Su1980} (SSH) model which possesses an arbitrary number of bands and underlies the results of the following sections. In Sec.~\ref{sec_dpts} we investigate the dynamical quantum phase transitions of this model, including their occurrence for quenches within a single phase and the role of the edge states. In Secs.~\ref{sec_ent} and \ref{sec_fidelity} we consider relations to the dynamical entanglement entropy and fidelity respectively. Finally in Sec.~\ref{sec_con} we conclude.

\section{A generalized Su-Schrieffer-Heeger model: A multiband topological insulator}
\label{sec_model}

In this paper we are interested in the behavior of one dimensional multiband topological insulators. To that end we introduce our analytically solvable model based on the SSH\cite{Su1980} model. The SSH model describes a chain with alternating hopping strengths, and hence with 2 atoms in the unit cell. Our generalisation to $N$ atoms in the  unit cell is
\begin{equation}\label{ham}
	H=\sum_{n}\Psi_{n}^\dagger\mathbf{M}_0\Psi_{n}+\left[\sum_{n}\Psi_{n+1}^\dagger\mathbf{M}_1\Psi_{n}+\textrm{H.c}\right]\,,
\end{equation}
where $\Psi_{n}^\dagger=(c_{n1}^\dagger,c_{n2}^\dagger,\ldots c_{nN}^\dagger)$ and $c_{nj}^\dagger$ is a fermionic creation operator in unit cell $n$ and site $j$. See Fig. \ref{figure_01} for a schematic of the model. The matrices are
\begin{equation}
	\left[\mathbf{M}_{0}\right]_{ij}=-J(1+g)\left(\delta_{i+1,j}+\delta_{i-1,j}\right)\,,
\end{equation}
and
\begin{equation}
	\left[\mathbf{M}_{1}\right]_{ij}=-J(1-g)\delta_{i+1,j}\,.
\end{equation}
$J$ is the hopping parameter and $g$ is the dimerisation, in the following we will scale energy and time such that $J=1$ and $\hbar$. For $N=2$ we recover the SSH model. This model possesses the same symmetries as the SSH model: particle-hole, `time-reversal', and their composite chiral symmetry; and hence is also in the symmetry class BDI within the ten-fold topological classification scheme.\cite{Ryu2010} See appendix \ref{app_symm} for more information. This class has a $\mathbb{Z}$ topological index and hence can possess many topologically protected edge states.

For periodic boundary conditions we can make the Fourier transform $\Psi_n=\frac{1}{\sqrt{N_w}}\sum_k\e^{\im k n}\Psi_k$ where $k=2\pi r/N_w$ with $r=1,2,\ldots N_w$ and $N_w$ is the number of unit cells. Then
\begin{eqnarray}
	H&=&\sum_{k}\Psi_k^\dagger\mathbf{M}_k\Psi_{k}\,,\textrm{ with}
	\nonumber\\
	\left[\mathbf{M}_{k}\right]_{ij}&=&\sigma_k\delta_{i+1,j}+\sigma_k^*\delta_{i-1,j}\,,
\end{eqnarray}
where
\begin{equation}
	\sigma_k=-(1+g)-(1-g)\e^{-\im k}\,.
\end{equation}
As $\mathbf{M}_{k}$ is a Toeplitz matrix this Hamiltonian can be solved exactly.\cite{Noschese2013} First we make the transformation $\Psi_k=\mathbf{U}_k^\dagger\tilde \Psi_k$ with $[\mathbf{U}_k]_{lm}=\delta_{lm}(\sigma_k/\sigma_k^*)^{\frac{l-\ell}{2}}=\delta_{lm}\e^{\im\phi_k(l-\ell)}$, where $\e^{\im\phi_k}=\sigma_k/|\sigma_k|$. The angle can be found from
\begin{equation}
	\tan\phi_k=\frac{(1-g)\sin k}{1+g+(1-g)\cos k}\,.
\end{equation}
$\ell$ is a constant which can be arbitrarily chosen, provided one uses a consistent scheme.\cite{Cooper2018} It causes a shift in the winding number, and a  convenient choice for us turns out to be $\ell=\frac{N-1}{2}$. We then find, with $\tilde{\mathbf{M}}_{k}=\mathbf{U}_k\mathbf{M}_{k}\mathbf{U}_k^\dagger$,
\begin{equation}
	\left[\tilde{\mathbf{M}}_{k}\right]_{ij}=|\sigma_k|\left(\delta_{i+1,j}+\delta_{i-1,j}\right)\,,
\end{equation}
which can be diagonalized with the transform
\begin{equation}
	\tilde c_{kj}=\sqrt{\frac{2}{N+1}}\sum_{\mu=1}^N\sin\left[\frac{\pi j\mu}{N+1}\right]\psi_{k\mu}
\end{equation}
where $\tilde\Psi_{k}^\dagger=(\tilde c_{k1}^\dagger,\tilde c_{k2}^\dagger,\ldots \tilde c_{kN}^\dagger)$.

Finally, with $\mu=1,2,\ldots N$, we have the eigenenergies
\begin{equation}
	\epsilon_{k\mu}=-2|\sigma_k|\cos\left[\frac{\mu\pi}{N+1}\right]\,,
\end{equation}
and
\begin{equation}\label{eigenstate}
	\psi_{k\mu}=\frac{1}{\sqrt{N_w}}
	\sqrt{\frac{2}{N+1}}\sum_{nj}\sin\left[\frac{\pi j\mu}{N+1}\right]\e^{\im\phi_k(j-\ell)-\im kj}c_{nj}\,.
\end{equation}
for the eigenstates.

This model has a critical gapless phase for $g=0$, and as we shall see for even $N$ it is topologically non-trivial if $g<0$. We further confine ourselves always to $|g|<1$ so that all hopping integrals are always positive. For odd $N$ the particle-hole symmetry imposes that there be a bulk band pinned exactly to zero energy. This can also be seen by considering the energy eigenstates with $\mu=(N+1)/2$. However, for odd $N$ we find no evidence of topologically protected edge states. For open boundary conditions no analytical solution exists for general $N$ and the system must be numerically diagonalized. A semi-analytical solution does however exist for the open $N=2$ chain.\cite{Shin1997}

\begin{figure}
\includegraphics*[width=0.8\linewidth]{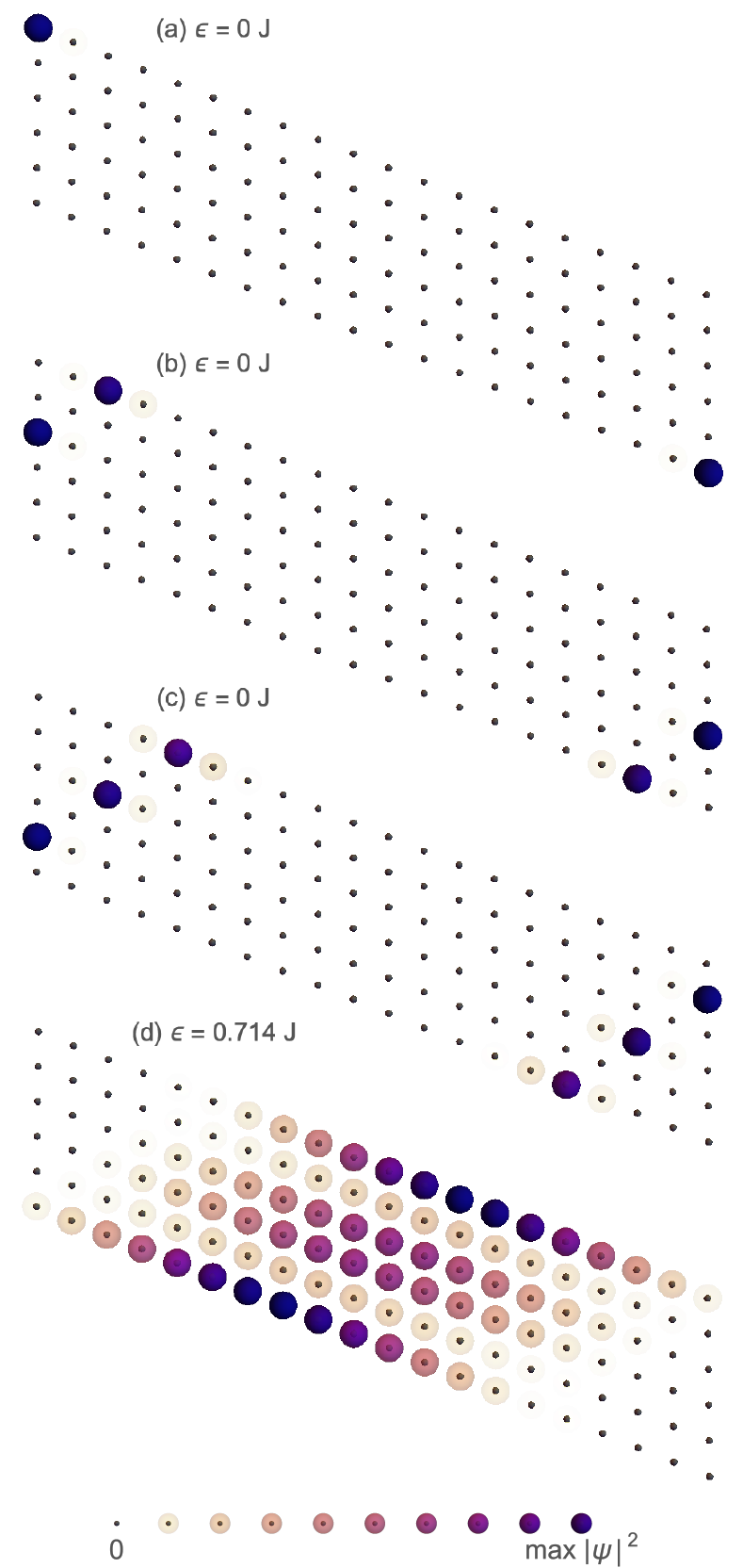}
\caption{Densities for eigenstates of the generalized SSH wire with $N=6$ and $g=-0.8$, i.e.~in its topologically non-trivial phase. The three positive energy ``zero energy'' edge states are shown. (a), (b), and (c), along with the lowest energy bulk state (d).}
\label{figure_02}
\end{figure}

In Fig.~\ref{figure_02} we show some exemplary densities of the eigenstates for $N=6$. Both edge states of the topologically non-trivial phase, and a lowest energy bulk state, are shown. The density distribution of the edge states forms a characteristic pattern on the edges, see also the densities for the case $N=4$ in Fig.~\ref{figure_02_4} in Appendix \ref{app_densities}.

\subsection{Topological phases and invariant}

In one dimension the topological invariant we are interested in is the winding number or Zak-Berry phase $\nu$.\cite{Berry1984,Zak1989,Ryu2006,Viyuela2014} The Zak-Berry phase for a single negative energy band $\mu$ is
\begin{equation}
	\nu_\mu=\frac{\varphi_\mu}{2\pi}=\im\int_0^{2\pi}\frac{\ud k}{2\pi}\langle u_{k\mu}|\partial_ku_{k\mu}\rangle\,,
\end{equation}
with the integral taken round the Brillouin zone and $|u_{k\mu}\rangle$ being the Bloch function of a negative energy band. This results in a $\mathbb{Z}$ invariant. The number of pairs of topologically protected edge states is then equal to the total winding number for all negative energy bands.\cite{Ryu2002}

\begin{figure}
\includegraphics*[width=0.8\linewidth]{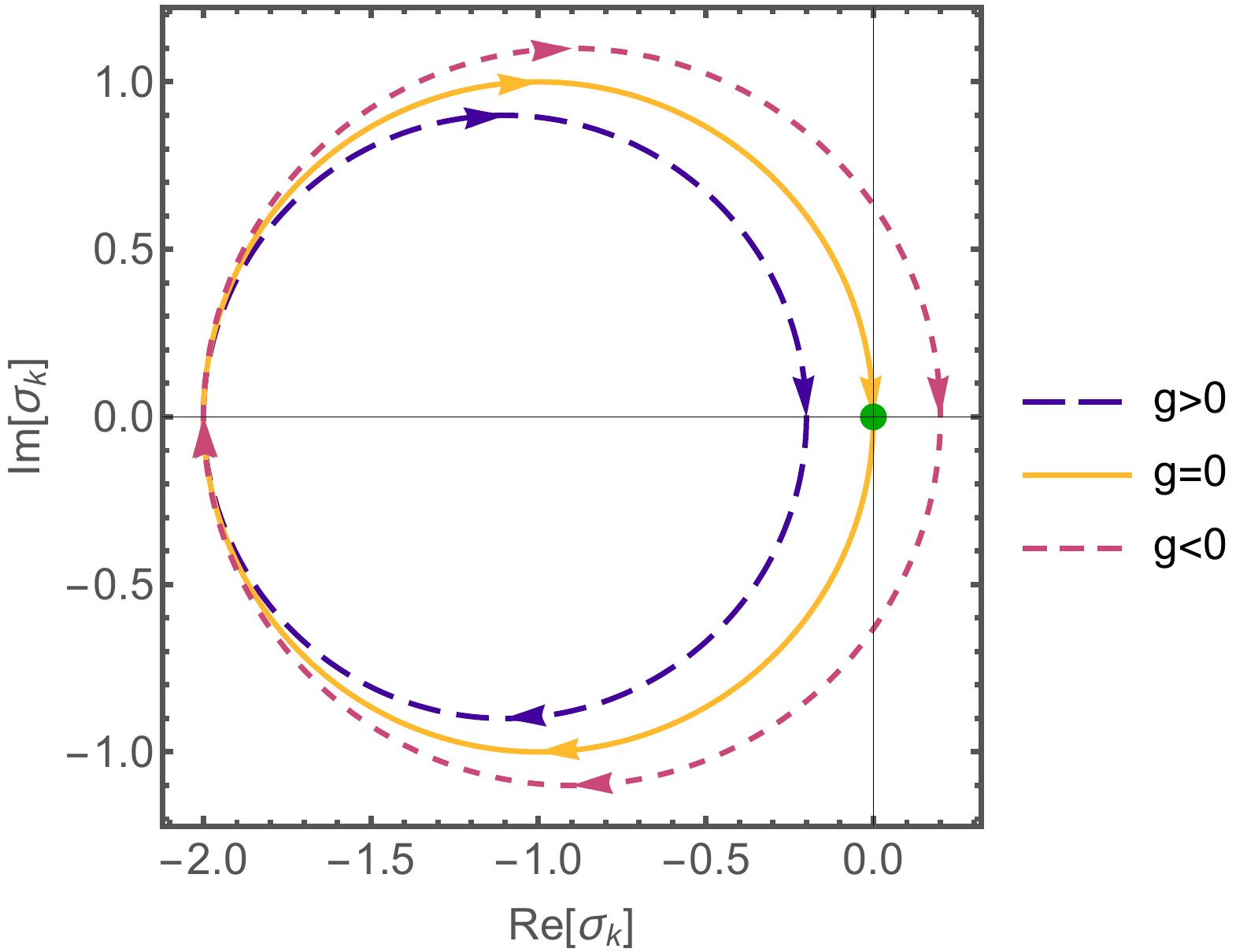}
\caption{Winding for the topological invariant, the loops show behaviour of $\sigma_k$ for $k:0\to2\pi$ for the system in the topologically non-trivial regime, $g<0$, and the trivial phase, $g>0$. At the critical point one can see that gap closes at $\sigma_{k=\pi}=0$, which distinguishes the two topological phases.}
\label{figure_03}
\end{figure}

Using the solution \eqref{eigenstate} we have
\begin{equation}
	\nu_\mu=-\frac{2}{N+1}\int_0^{2\pi}\frac{\ud k}{2\pi}\phi_k'\sum_{j}(j-\ell)\sin^2\left[\frac{\pi j\mu}{N+1}\right]\,.
\end{equation}
By considering the behaviour of the phase $\phi_k$, see Fig.~\ref{figure_03}, we can see that
\begin{equation}
	\nu_\mu=\hside(-g)\frac{2}{N+1}\sum_{j}(j-\ell)\sin^2\left[\frac{\pi j\mu}{N+1}\right]\,,
\end{equation}
where $\hside(g)$ is the Heaviside theta function. With the help of expressions from App.~\ref{app_exp} we have $\nu_\mu=\hside(-g)$, and summing over all negative energy bands we find, for even $N$,
\begin{equation}
	\nu=\sum_{\mu:\epsilon_{k\mu}<0}\nu_\mu=\frac{N}{2}\hside(-g)\,.
\end{equation}
Thus in a topologically non-trivial phase with invariant $\nu=N/2$ we have $N/2$ pairs of protected edge modes at the ends of the chain. Therefore a chain with unit cell size $N$ always has $N$ topologically protected edge modes in its topologically non-trivial phase.

For odd $N$ one would need to consider whether to include the zero-energy flat band. However the system is not gapped at zero energy due to the presence of this flat band. In this sense it can not be regarded as a topological insulator. Furthermore we find no evidence of topologically protected edge states for the odd $N$ chains. We therefore do not consider its invariant here.

\subsection{Density of states}

\begin{figure}
\includegraphics*[width=0.99\linewidth]{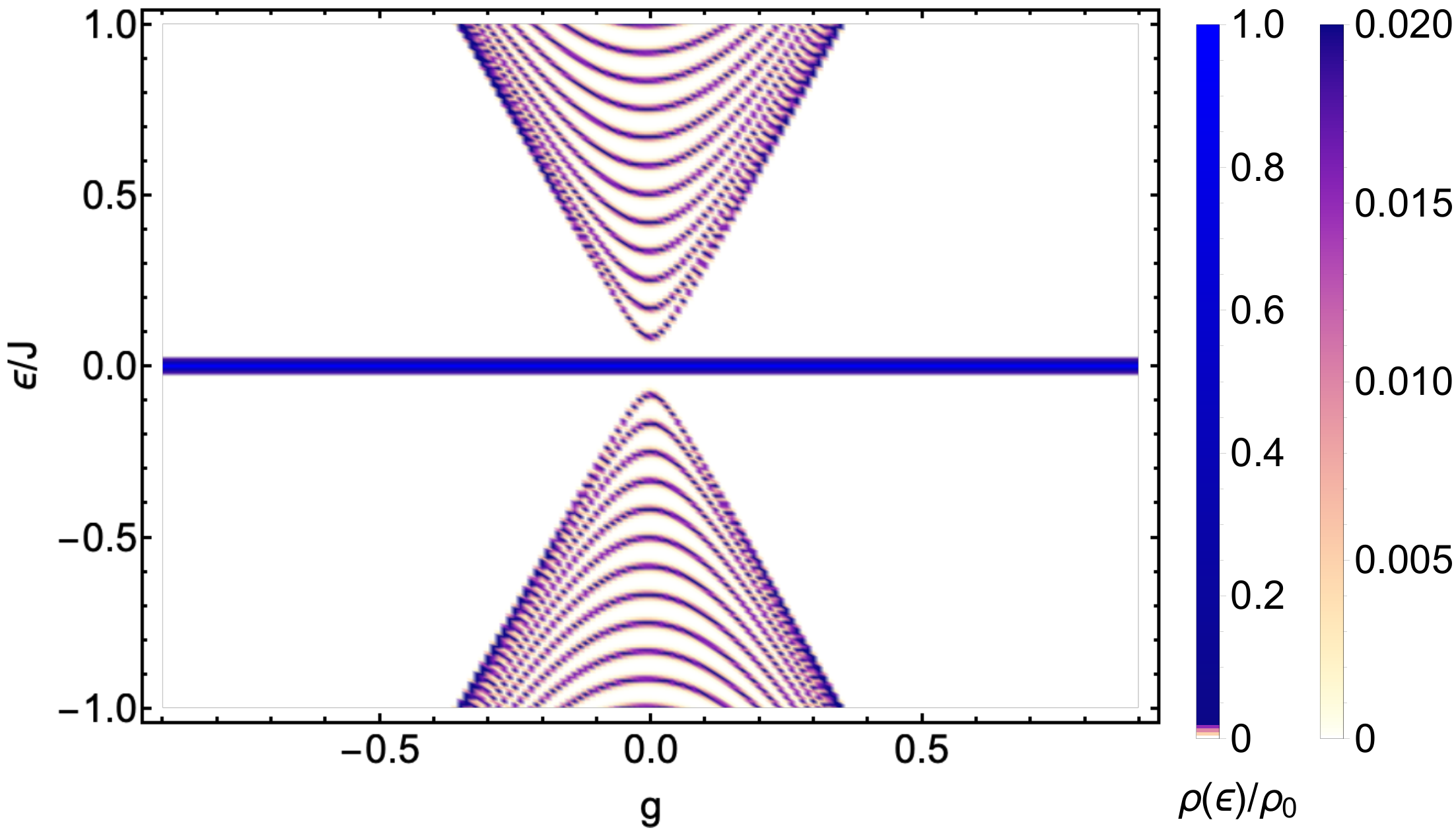}
\caption{The density of states as a function fo dimerisation for a chain with $N=3$. The bulk band pinned to zero energy is clearly visible. The right hand scale is a zoom in on the total scale, and $\rho_0=2821J^{-1}$.}
\label{figure_04}
\end{figure}

\begin{figure}
\includegraphics*[width=0.99\linewidth]{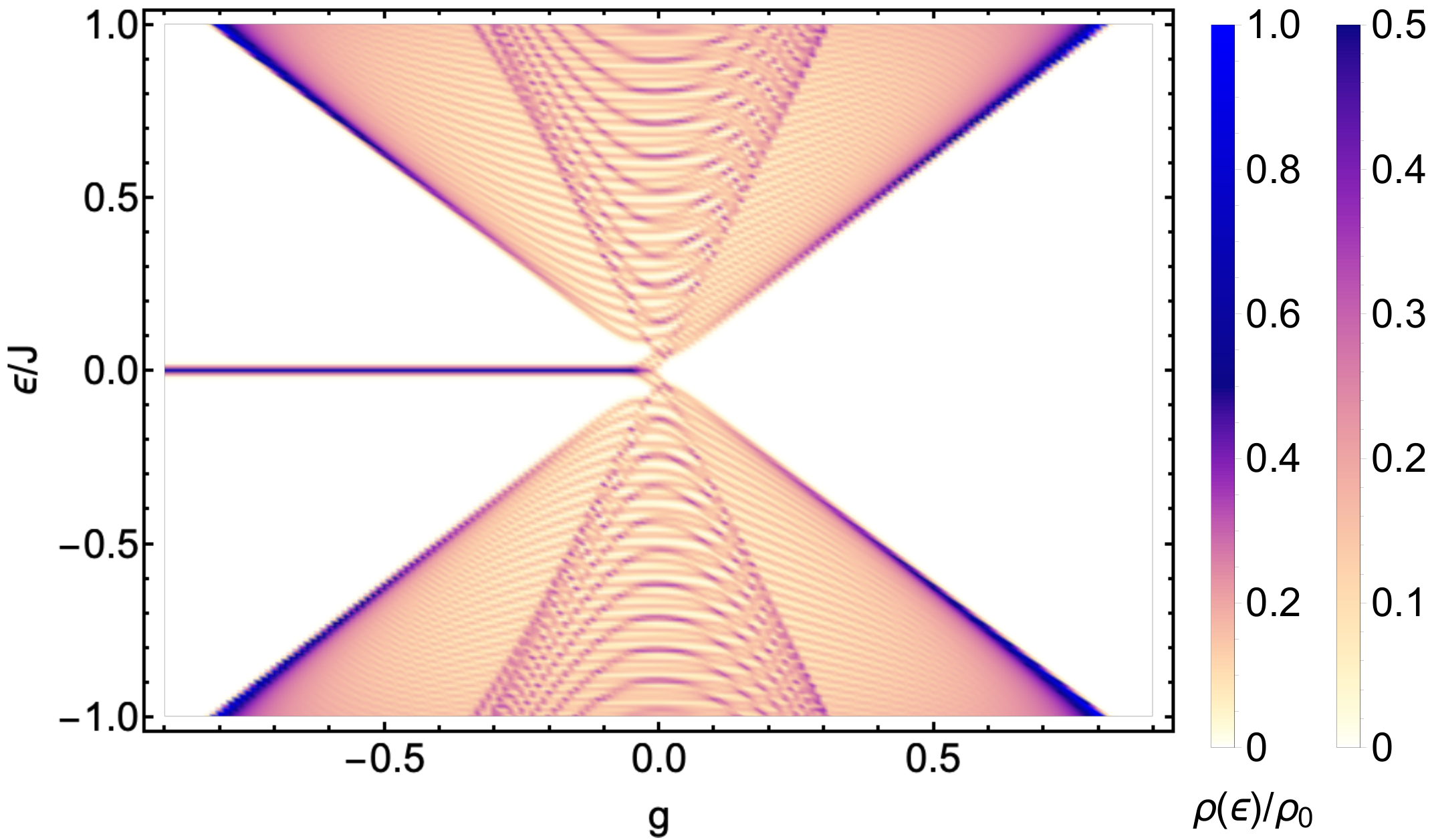}
\caption{The density of states as a function fo dimerisation for a chain with $N=4$. The right hand scale is z zoom in on the total scale, and $\rho_0=339.1J^{-1}$.}
\label{figure_05}
\end{figure}

In Figs.~\ref{figure_04} and \ref{figure_05} we show the density of states for $N=3$ and $N=4$ as a function of the dimerization. A continuum density of states $\rho(\epsilon)$ has been obtained from the finite system size data by the following broadening procedure:
\begin{equation}
	\rho(\epsilon)=\frac{1}{\epsilon_0\sqrt{\pi}}\sum_n\e^{-\frac{(\epsilon-\epsilon_n)^2}{\epsilon_0^2}}
\end{equation}
with $\epsilon_0=0.01t$. For odd $N$, Fig.~\ref{figure_04}, the bulk zero-energy bands are clearly visible, along with the closing and opening of the gap at the critical $g=0$. For even N, Fig.~\ref{figure_05}, the topological phase transition can be seen with the zero energy edge sates appearing for $g<0$ and the gap closing at $g=0$.

\section{Dynamical quantum phase transitions}
\label{sec_dpts}

A dynamical quantum phase transition is said to occur when the Loschmidt amplitude becomes zero. The Loschmidt amplitude, which is the overlap between an initial state $|\psi_1\rangle$ and its counterpart time evolved by a Hamiltonian $H_2$, $\e^{-\im H_2t}|\psi_1\rangle$, is
\begin{equation}
	L(t)=\left\langle\psi_1\left|\e^{-\im H_2t}\right|\psi_1\right\rangle\,.
\end{equation}
For $\im t\to z$, with $z$ a complex variable, the Loschmidt amplitude is a generalized partition function with Fisher zeros\cite{Fisher1965,Heyl2013} in the complex plane where $L(-\im z)=0$. When these zeros cross the imaginary axis we have a dynamical quantum phase transition. Zeros in the Loschmidt amplitude correspond to the appearance of cusps in the corresponding return rate
\begin{equation}\label{returnrate}
	l(t)=-\frac{1}{NN_w}\ln |L(t)|\,,
\end{equation}
which is introduced in analogy with the normal free energy. In our case the initial state $|\psi_1\rangle$ is taken to be the ground state of the Hamiltonian \eqref{ham} with dimerisation $g_1$ at half-filling. Time evolution happens with the Hamiltonian $H_2$, which has dimerisation $g_2$.

\subsection{Bulk contributions}

Firstly in order to find an expression for the bulk Loschmidt amplitude in the thermodynamic limit for our multiband model, we start from the expression\cite{Levitov1996,Klich2003,Rossini2007,Sedlmayr2018}
\begin{equation}
	\label{rle}
	L(t)=\Det\mathbf{M}\equiv\Det\left[1-\mathbf{\C}+\mathbf{\C}\e^{\im {\bm H}_12t}\right]\,,
\end{equation}
where $\mathbf{\C}$ is the correlation matrix for the initial state $\C_{ij}=\langle\psi_1|\hat{c}^\dagger_i\hat{c}_j|\psi_1\rangle$. The vectors $\{c_j\}$ form a basis of the Hilbert space and ${\bm H}_{1,2}$ is $H_{1,2}$ written in this basis. Let us write the matrix $\mathbf{M}$ in the eigenbasis of $H_2$, then ${\bm H}_2$ is diagonal and $\mathbf{\C}$ and $\mathbf{M}$ become block diagonal with each block being labelled by momentum $k$. We then find
\begin{equation}
	\label{rle2}
	L(t)=\prod_k\det\mathbf{M}_k\equiv\prod_k\det\left[1-\mathbf{\C}^k+\mathbf{\C}^k\e^{\im {\bm H}_2^kt}\right]\,.
\end{equation}
We use lowercase $\det$ for the determinant over the subspace labelled by the $\lambda$ quantum numbers of the eigenstates. We also introduce the eigenvalues $\lambda_i(t)$ of the Loschmidt matrix\cite{Sedlmayr2018} $\mathbf{M}(t)$ such that
\begin{equation}
	L(t)\equiv\prod_i\lambda_i(t)\,.
\end{equation}
 A dynamical quantum phase transition is caused by one of these eigenvalues becoming zero. Note that for the case of half-filling which we focus on here, where the negative energy bands are full and the positive energy bands are empty, half of the eigenvalues $\lambda_i(t)$ are identical to $1$.

The correlation matrix can be expressed as
\begin{eqnarray}\label{corrmatrix}
	\mathbf{\C}^k_{\lambda\lambda'}&=&\left(\frac{2}{N+1}\right)^2\sum_{ij}\sum_{\mu:\epsilon_{k\mu}<0}
	\sin\left[\frac{\pi\lambda i}{N+1}\right]\sin\left[\frac{\pi\mu i}{N+1}\right]\nonumber\\&&
	\sin\left[\frac{\pi\lambda' j}{N+1}\right]\sin\left[\frac{\pi\mu j}{N+1}\right]\e^{\im\delta\phi_k(i-j)}\,.
\end{eqnarray}
We have introduced the ``angle'' between the initial and time-evolving Hamiltonians $\delta\phi_k=\phi_k(g_2)-\phi_k(g_1)$. These sums can in principle be performed analytically, but the resulting expression is not in general very compact, and we exclude giving it here explicitly. The result of the sum over $\mu$ is given by Eq.~\eqref{musum} in Appendix \ref{app_exp}.

For $N=2$ this returns the standard result\cite{Vajna2015}
\begin{equation}
	\det\mathbf{M}_k=\cos[\tau_k]-\im\cos[\delta\phi_k] \sin[\tau_k]\,,
\end{equation}
where we have defined a scaled time for each momenta:
\begin{equation}
	\tau_k=t|\sigma_{k}(g_1)|\,.
\end{equation}
As can be readily seen this only has zeroes if
\begin{equation}
\label{cos_0}
\cos[\epsilon_{k_c2} t]=0 
\end{equation}
can be satisfied. In that case the critical times are
\begin{equation}
	t_{cn}=\frac{\pi(2n+1)}{2|\sigma_{k_c}(g_1)|}\,,
\end{equation}
for $n\in\mathbb{Z}$. We will define a critical timescale $t_c$ as the smallest critical time.

The expression for the angle does not depend on $N$ and is, as for the SSH model,
\begin{eqnarray}
	\cos[\delta\phi_k]&=&\frac{(1-g_1g_0)\cos k-1-g_1g_0}{\sqrt{1+g_1^2+(g_1^2-1)\cos k}}
	\nonumber\\&&\quad\times
	\frac{1}{\sqrt{1+g_0^2+(g_0^2-1)\cos k}}\,,
\end{eqnarray}
which for symmetric quenches $g_0=-g_1$ reduces to
\begin{equation}
	\cos[\delta\phi_k]=\frac{g_0^2-1+(1+g_0^2)\cos k}{1+g_0^2+(g_0^2-1)\cos k}\,.
\end{equation}
Solving $\cos[\delta\phi_{k_c}]=0$ defines the critical momentum
\begin{equation}
	k_c=\cos^{-1}\left[\frac{1+g_1g_0}{1-g_1g_0}\right]\,,
\end{equation}
which only has real solutions for $g_1g_2<0$.\cite{Vajna2015} I.e.~for the case when the quench crosses the topological phase transition. For systems with more structure in the Brillouin zone, such as the long range Kitaev chain, this can have multiple solutions.\cite{Sedlmayr2018}

For $N=3$ one finds
\begin{equation}
	\det\mathbf{M}_k=\left[\cos\frac{\tau_k}{\sqrt{2}}-\im\cos[\delta\phi_k] \sin\frac{\tau_k}{\sqrt{2}}\right]^2\,,
\end{equation}
independent of whether the flat band at zero energy is completely full or empty. This has the same behaviour as $N=2$, and in the following we will focus purely on even $N$ chains.

For $N=4$ we find
  \begin{multline}
    \label{det4}
  \det \mathbf{M}_k(t)=\\
  \textrm{c}_4\left(\tau_{k},\cos[\delta\phi_k]\right)+\im \,\cos[\delta\phi_k]\,\textrm{s}_4\left(\tau_{k},\cos[\delta\phi_k]\right)\,,
\end{multline}
where
\begin{multline}
  \textrm{c}_4(\tau,u)=
\frac{8}{25} \left(1 - u^2\right)^2 + \frac{\cos[\tau]}{10} \left(5 - u^2 - 4 u^4\right)\\+\frac{\cos[\sqrt{5} \tau]}{50} \left(9 + 37 u^2 + 4 u^4\right)\,,
\end{multline}
and
\begin{multline}
  \textrm{s}_4(\tau,u)=\\
  -\frac{2\sin[\tau]}{\sqrt{5}}\left(1-u^2\right)-\frac{\sin[\sqrt{5}\tau]}{5}\left(3+2u^2\right)\,.
\end{multline}
This also has solutions for the critical times at the same critical momenta $k_c$ as for $N=2$ and then, using Eq.~\eqref{cos_0}, we find
\begin{equation}\label{n4crit}
	\det\mathbf{M}_{k_c}=1+\frac{3}{5}\cos[\tau_c]+\frac{2}{25}\cos[\sqrt{5}\tau_c]\,,
\end{equation}
with
\begin{equation}
	\tau_c\equiv\tau_{k_c}=t|\sigma_{k_c}(g_1)|\,.
\end{equation}
As the ratio of 1 and $\sqrt{5}$ is not rational the solutions of $\det\mathbf{M}_{k_c}=0$ with the determinant from Eq.~\eqref{n4crit} give quasiperiodic critical times\cite{Karrasch2013}, rather than the periodic solutions as for the SSH model and other simple two-band topological systems. They can be determined numerically.
However these are not the only solutions.

An additional class of solutions also exist for $\cos[\delta\phi_k]\neq0$. From $\Im\det\mathbf{M}_k=0$, where $\Im$ refers to the imaginary part, we find
\begin{equation}\label{cossq}
	\cos^2[\delta\phi_k]=\frac{2\sqrt{5}\sin[\tau_k]+3\sin[\sqrt{5}\tau_k]}{2\sqrt{5}\sin[\tau_k]-2\sin[\sqrt{5}\tau_k]}\equiv f[\tau_k]\,.
\end{equation}
Substituting this into $\Re\det\mathbf{M}_k=0$, with $\Re$ the real part, and solving it gives a complete list of zeros when combined with the solutions at the critical momentum. These additional solutions can occur at arbitrary momenta, and lead to aperiodic critical times. As these additional topological phase transitions do not require $\cos[\delta\phi_k]=0$ one could imagine it is possible that they can also occur for quenches which do not cross the topological phase boundary, and indeed, as we show in the next section this is now possible.

One can also derive an analytical expression for $N=6$, though it is too lengthy to place here. However, $\det\mathbf{M}_{k}$ for $N=6$ can still be written in a form like Eq.\eqref{det4}, with $\textrm{c}_4\left(\tau_{k},\cos[\delta\phi_k]\right)\to\textrm{c}_6\left(\tau_{k},\cos[\delta\phi_k]\right)$ and $\textrm{s}_4\left(\tau_{k},\cos[\delta\phi_k]\right)\to\textrm{s}_6\left(\tau_{k},\cos[\delta\phi_k]\right)$.
%\begin{multline}
%\label{det6}
%\det \mathbf{M}_k(t) = \\
%\textrm{c}_6\left(\tau_{k},\cos[\delta\phi_k]\right)+\im \,\cos[\delta\phi_k]\,\textrm{s}_6\left(\tau_{k},\cos[\delta\phi_k]\right)\,,
%\end{multline}
As before $\textrm{c}_6$ and $\textrm{s}_6$ are polynomials in $\e^{\im\tau_{k}}$, and in the second argument $\cos[\delta\phi_k]$. Thus, all observations made for $N=4$ are qualitatively similar for $N=6$. For larger $N$ the expressions become rapidly unwieldy.

\begin{figure}
\includegraphics*[width=0.99\columnwidth]{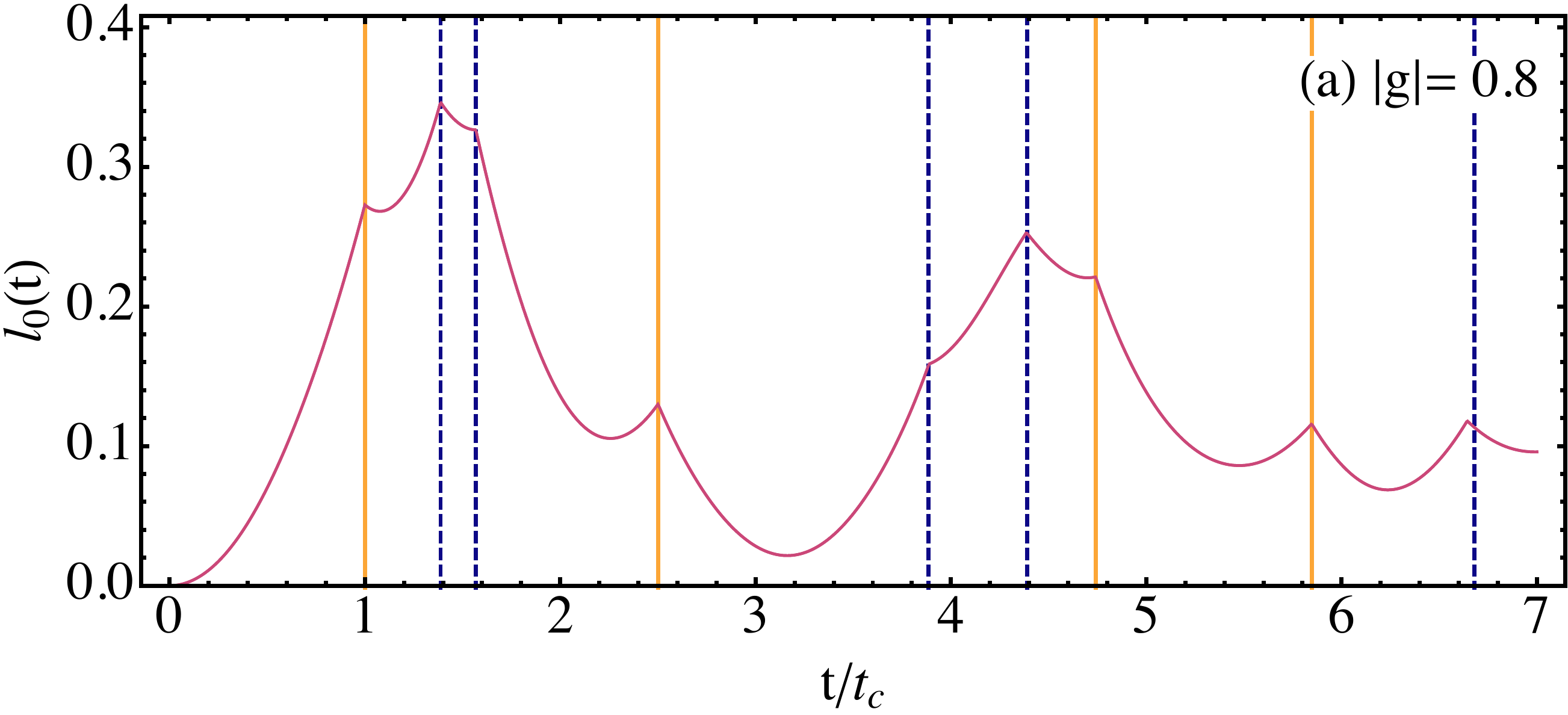}\\
\includegraphics*[width=0.99\columnwidth]{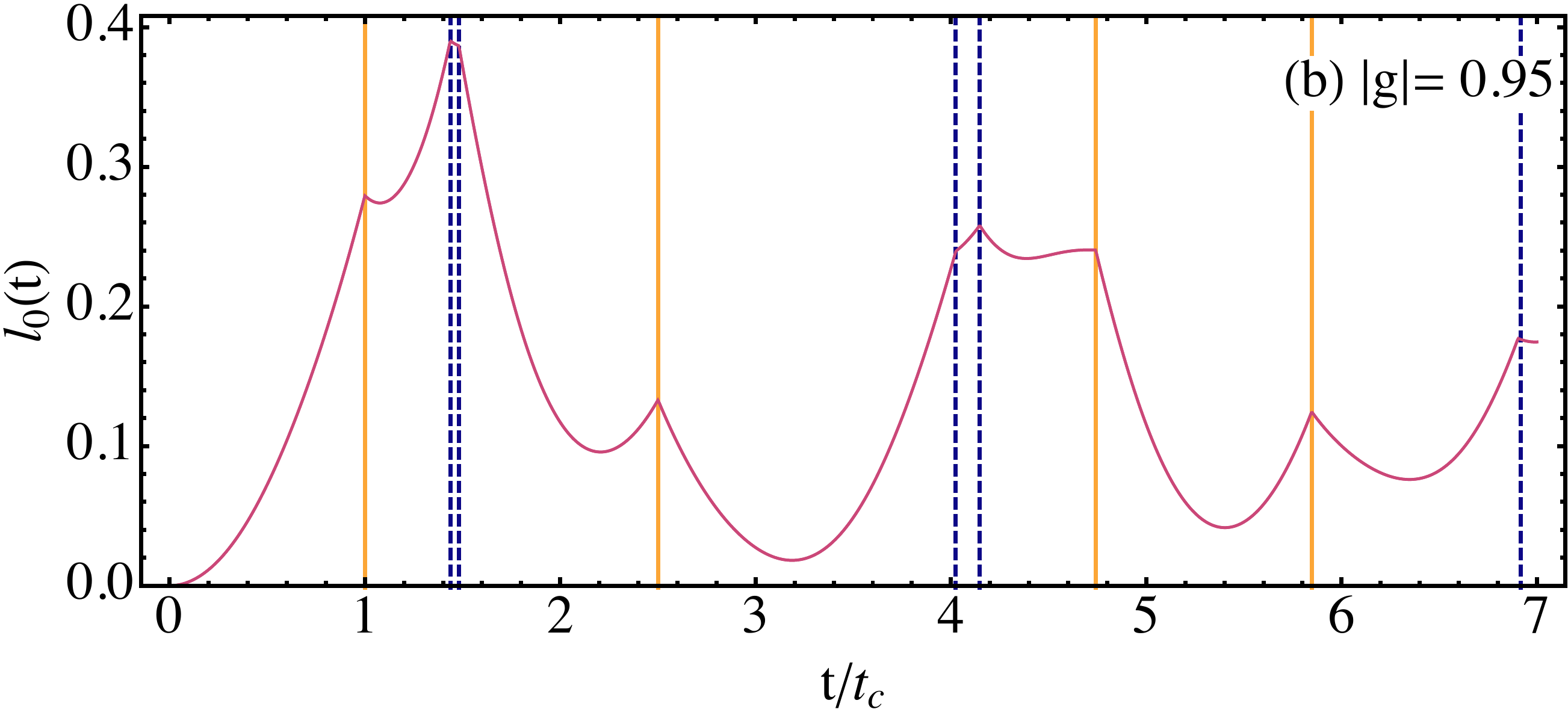}
\caption{The bulk return rate $l_0(t)$ for $N=4$. For (a) $g_1=-g_2=0.80$, and (b) $g_1=-g_2=0.95$. The orange lines show the location of the dynamical quantum phase transitions calculated for the critical momenta and the (dashed) purple lines for $\cos[\delta\phi_k]\neq0$.}
\label{figure_06}
\end{figure}

\begin{figure}
\includegraphics*[width=0.99\columnwidth]{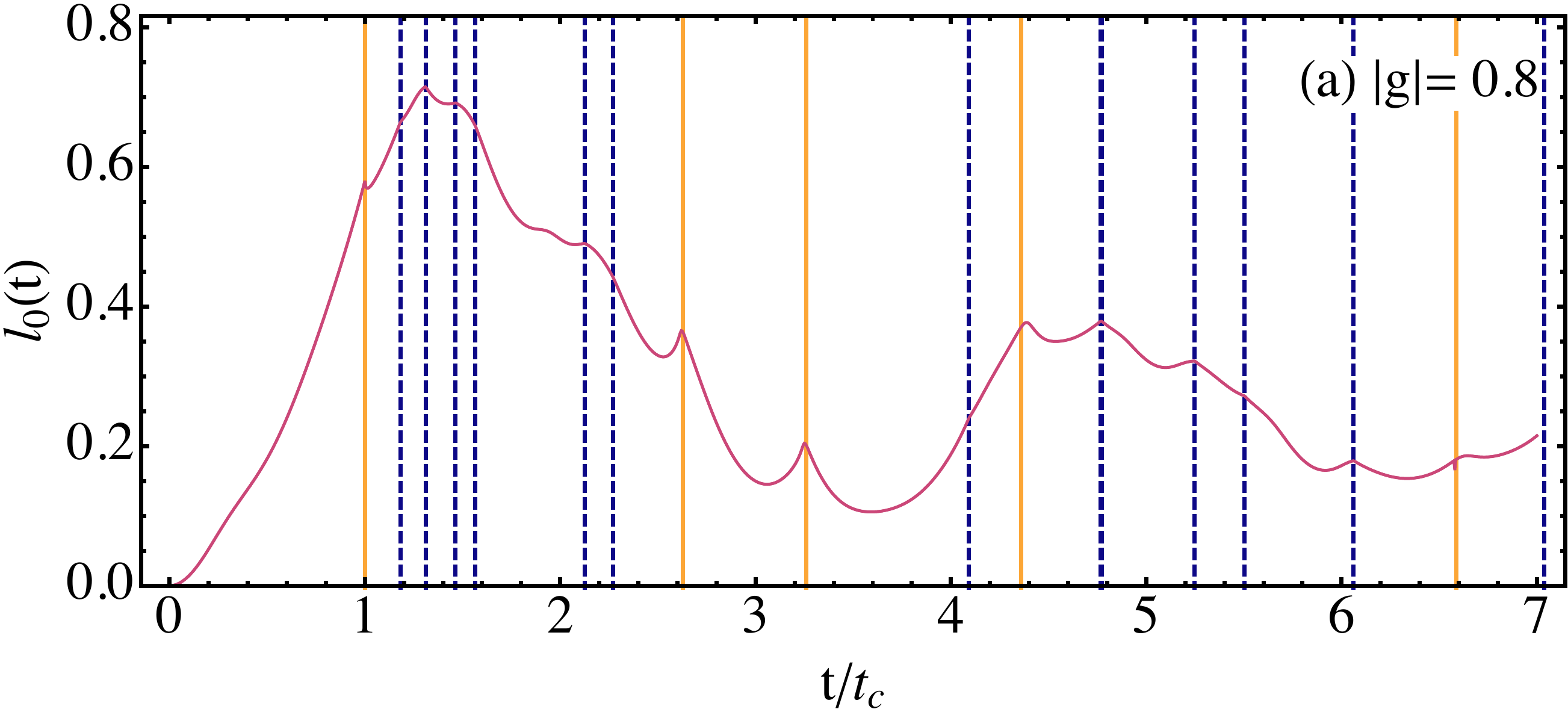}\\
\includegraphics*[width=0.99\columnwidth]{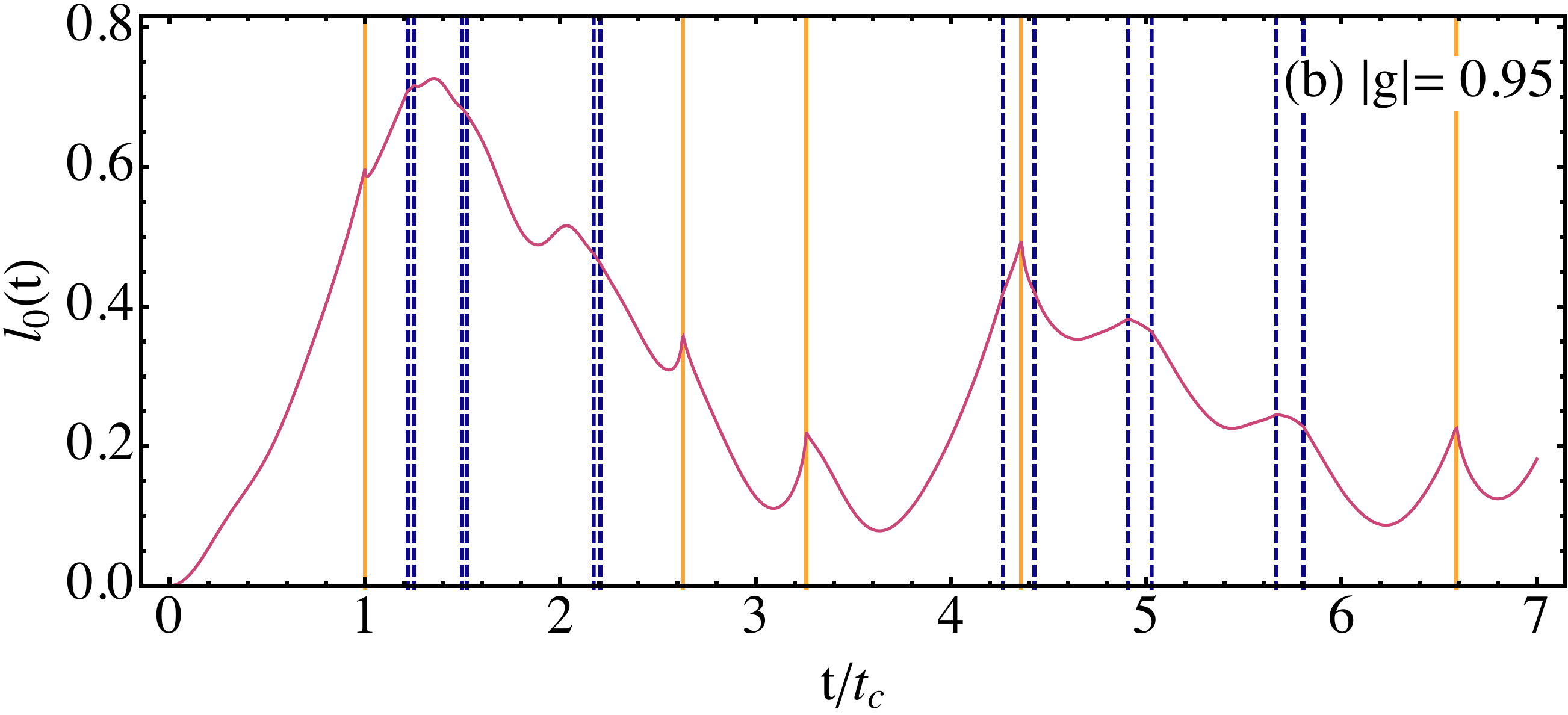}
\caption{The bulk return rate $l_0(t)$ for $N=6$. For (a) $g_1=-g_2=0.80$, and (b) $g_1=-g_2=0.95$. The orange lines show the location of the dynamical quantum phase transitions calculated for the critical momenta and the (dashed) purple lines for $\cos[\delta\phi_k]\neq0$.}
\label{figure_07}
\end{figure}

As we are interested also in the effects of the edge states\cite{Sirker2014,Sedlmayr2018} on the return rate we define, for $N_w\to\infty$,
\begin{equation}\label{scaling}
	l(t) \sim l_{0}(t)+\frac{l_{B}(t)}{N_w}\,.
\end{equation}
$l_0(t)$ is the bulk return rate in the limit $N_w\to\infty$, and $l_B(t)$ the boundary correction. In the limit $N_w\to\infty$ the return rate can be calculated as an integral over $k$. In Figs.~\ref{figure_06} and \ref{figure_07} $l_0(t)$ is shown for the symmetric quenches $g\to-g$ where $|g|=0.8$ and $|g|=0.95$ for both $N=4$ and $N=6$. For the bulk return rate the result is independent of whether the quench would be from $|g|\to-|g|$ or from $-|g|\to|g|$ .

\subsection{Dynamical quantum phases transitions for quenches within a single topological phase}

\begin{figure}
\includegraphics*[width=0.99\columnwidth]{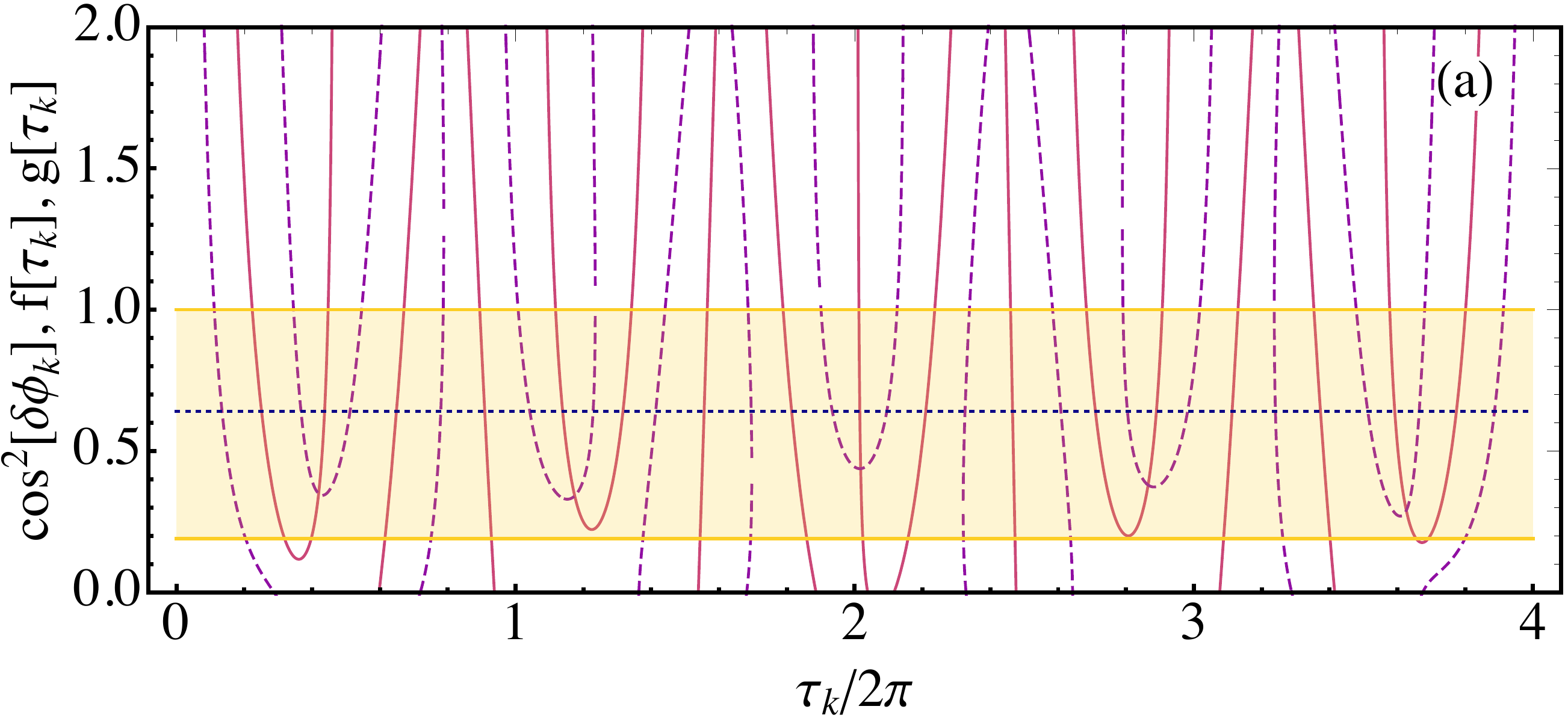}\\
\includegraphics*[width=0.99\columnwidth]{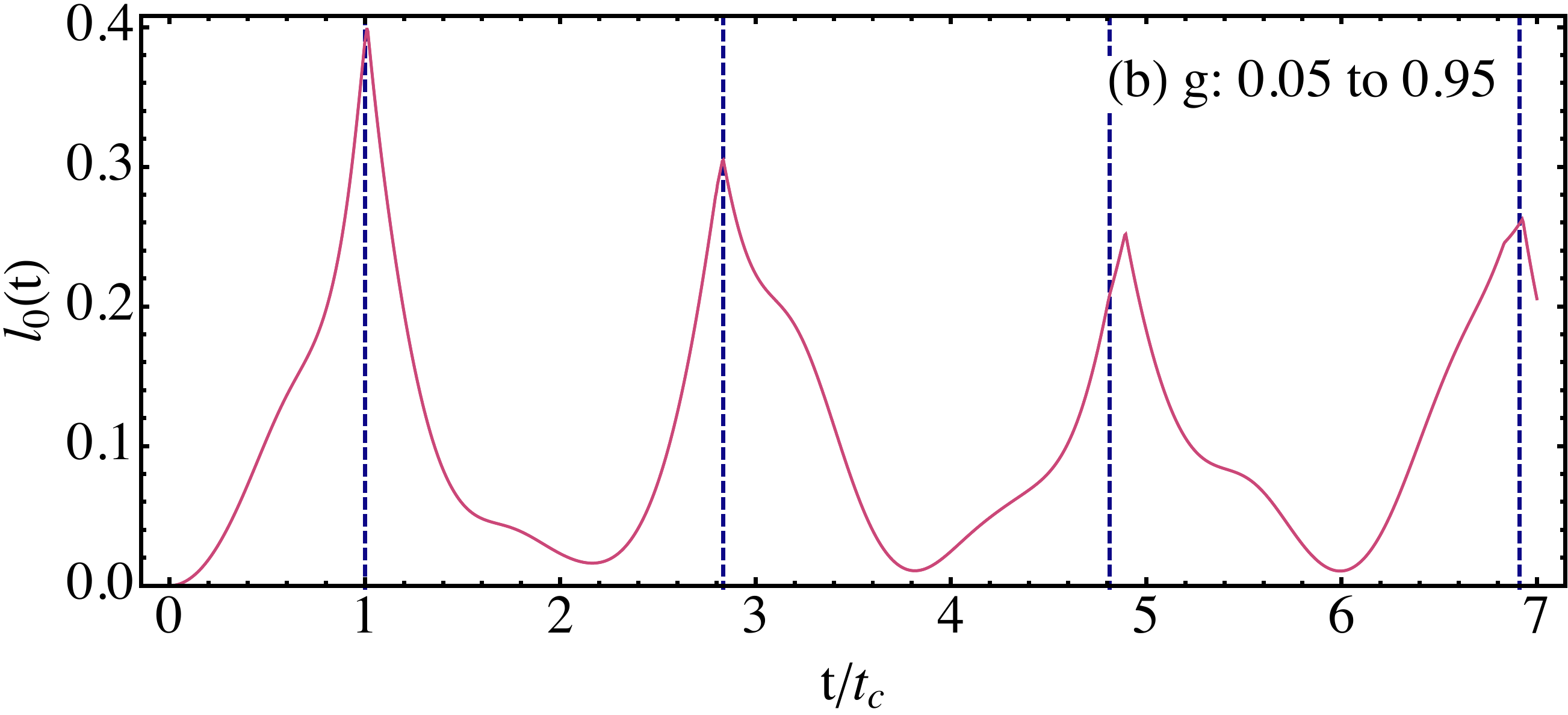}
\caption{In panel (a) we show the conditions for a dynamical quantum phase transition when $\cos[\delta\phi_k]\neq0$. Solutions exist where $f(\tau)$, the solid line, and $g(\tau)$, the dashed line, cross within the shaded region. The shaded region represents the possible values of $\cos^2[\delta\phi_k]$ for $g_1=0.05$ and $g_2=0.95$. If instead $g_1=0.2$ then the lower limit of $\cos^2[\delta\phi_k]$, and hence the shaded region, is the dotted line. Note that in that case there are no longer any critical times for this time range. In panel (b) the bulk return rate for $N=4$ where $g_1=0.05$ and $g_2=0.95$ is shown. The (dashed) purple lines show the critical times for $\cos[\delta\phi_k]\neq0$.}
\label{figure_08}
\end{figure}

To have a dynamical quantum phase transition one requires that
\begin{equation}
  \label{re_im}
	\Re\det\mathbf{M}_k=0\textrm{ and }\Im\det\mathbf{M}_k=0\,.
\end{equation}
The imaginary part of the determinant is always zero for the critical momentum, but these occur only for quenches across the topological phase transition in this model. If we wish to find a dynamical quantum phase transition for a quench within a phase we must consider the alternative solutions for $\cos[\delta\phi_k]\neq0$. A geometrical solution to Eq.\eqref{re_im} is plotted in Fig.\ref{figure_08}(a). Rewriting $\Re\det\mathbf{M}_k=0$ as $\cos^2[\delta\phi_k]\equiv g[\tau_k]$ then the solutions to $\det\mathbf{M}_k=0$ can be rewritten as $f[\tau_k]=g[\tau_k]$ with the constraint $0\leq k<2\pi$. In Fig.\ref{figure_08}(a) we plot $f[\tau_k]$ and $g[\tau_k]$, critical times for the dynamical quantum phase transitions will occur when they cross in the range which corresponds to the condition $0\leq k<2\pi$, the shaded region. This region corresponds to the possible values of $\cos^2[\delta\phi_{k}]$ which ranges from 1 at $k=0$ down to, for the case $g_1g_2>0$,
\begin{equation}
	\cos^2[\delta\phi_{\tilde{k}}]\equiv\frac{4g_1g_2}{g_1+g_2}\,.
\end{equation}
Note that the minimum of $\cos^2[\delta\phi_{k}]$ for $g_1g_2<0$ is always 0. If $g_1$ and $g_2$ are sufficiently far from each other then this still has solutions, see Fig.~\ref{figure_08}, and therefore dynamical quantum phase transitions can still occur.

\subsection{Boundary contributions to the return rate}

In the thermodynamic limit $N_w\to\infty$ there is no longer any effect from the boundary on the return rate or Loschmidt amplitude and we must consider the leading $1/N$ term $l_B(t)$. The boundary contribution can show very asymmetric behaviour for quenches across the topological phase transition in the two different directions from trivial to non-trivial and vice-versa. For two band topological insulators and superconductors it was found that the principle contribution to $l_B(t)$ for quenches into the topologically non-trivial phase originates in particular from several eigenvalues $\lambda_i(t)$ which become zero and stay zero in-between dynamical quantum phase transitions. The number of eigenvalues appeared to be related to the number of edge states, and by extension to the bulk topological invariants.\cite{Sedlmayr2018} This causes plateaus in $l_B(t)$ which are absent in the opposite direction a quench. A form of dynamical bulk-boundary correspondence.

\begin{figure}
\includegraphics*[width=0.99\columnwidth]{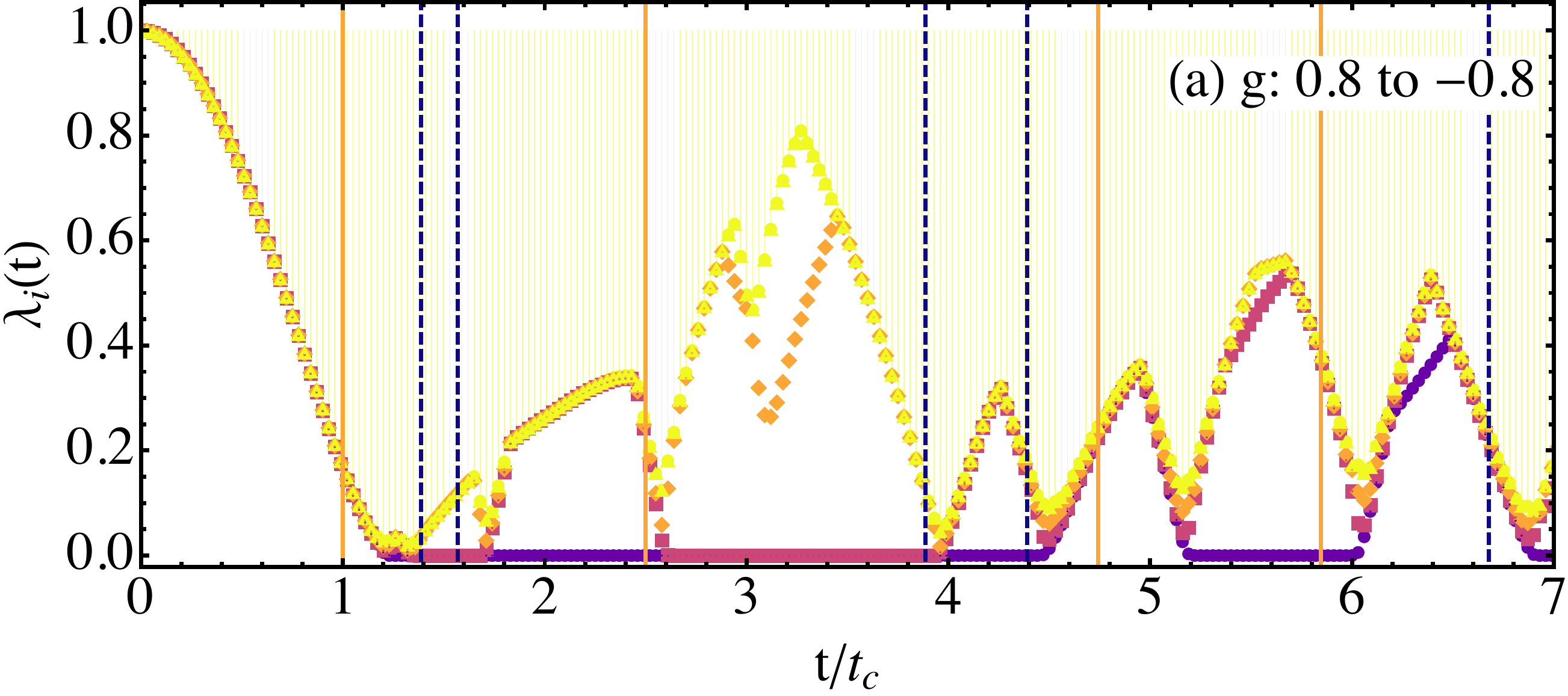}\\
\includegraphics*[width=0.99\columnwidth]{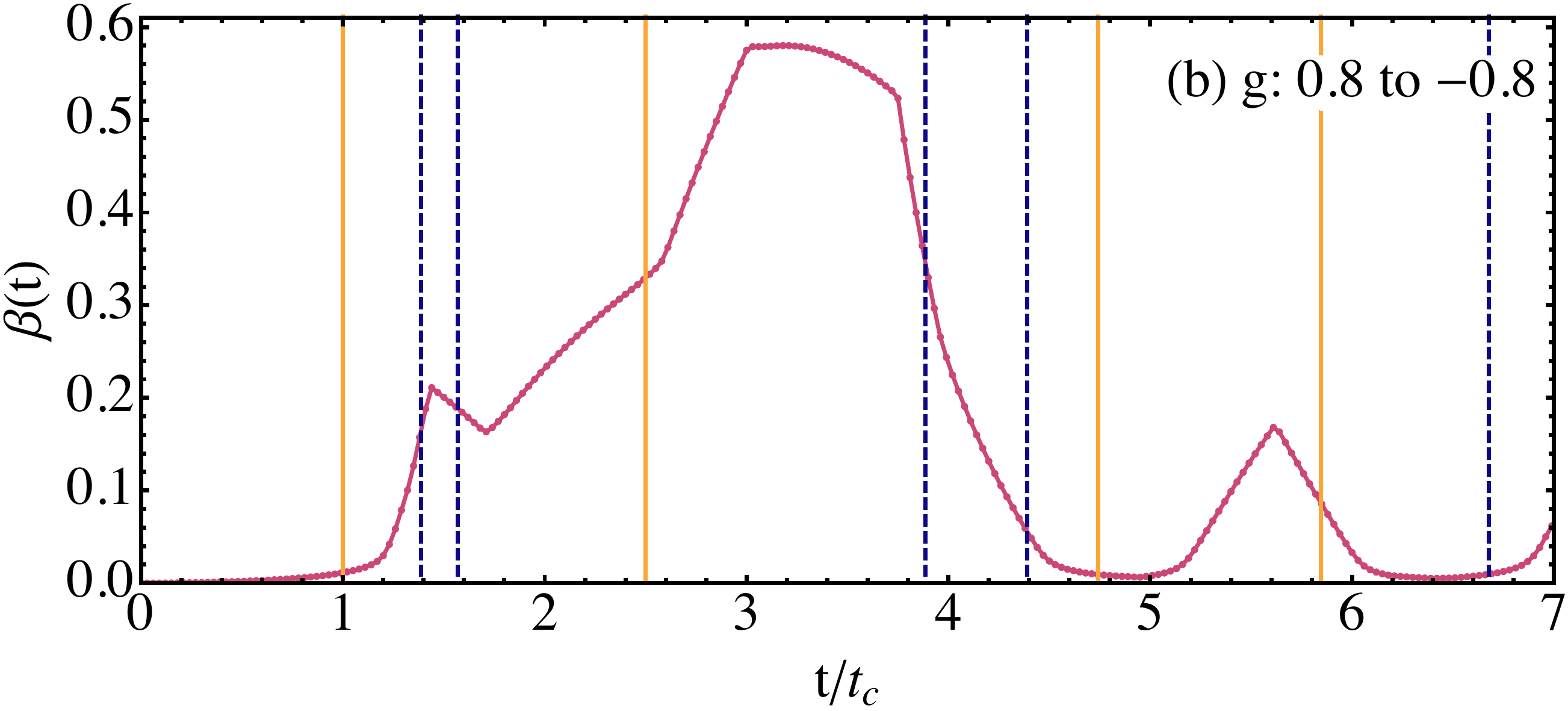}
\caption{(a) The smallest eigenvalues of $\mathbf{M}$, $\lambda_i(t)$, and (b) $\beta(t)$, for $N=4$, $N_w=160$, and $g_1=-g_2=0.8$ (trivial to non-trivial quench). For open boundary conditions. The orange lines show the location of the dynamical quantum phase transitions calculated for $\cos[\delta\phi_k]=0$ and (dashed) purple for $\cos[\delta\phi_k]\neq0$.}
\label{figure_09}
\end{figure}

\begin{figure}
\includegraphics*[width=0.99\columnwidth]{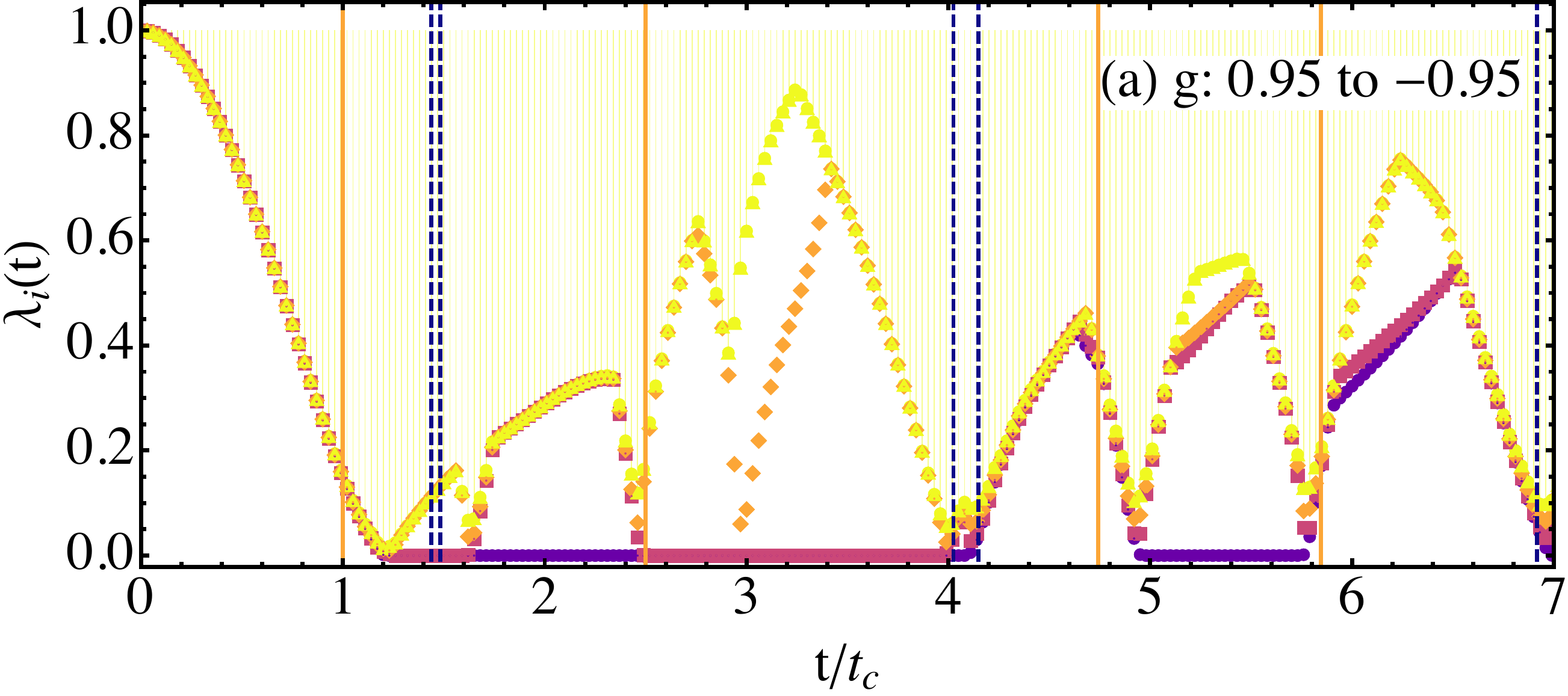}\\
\includegraphics*[width=0.99\columnwidth]{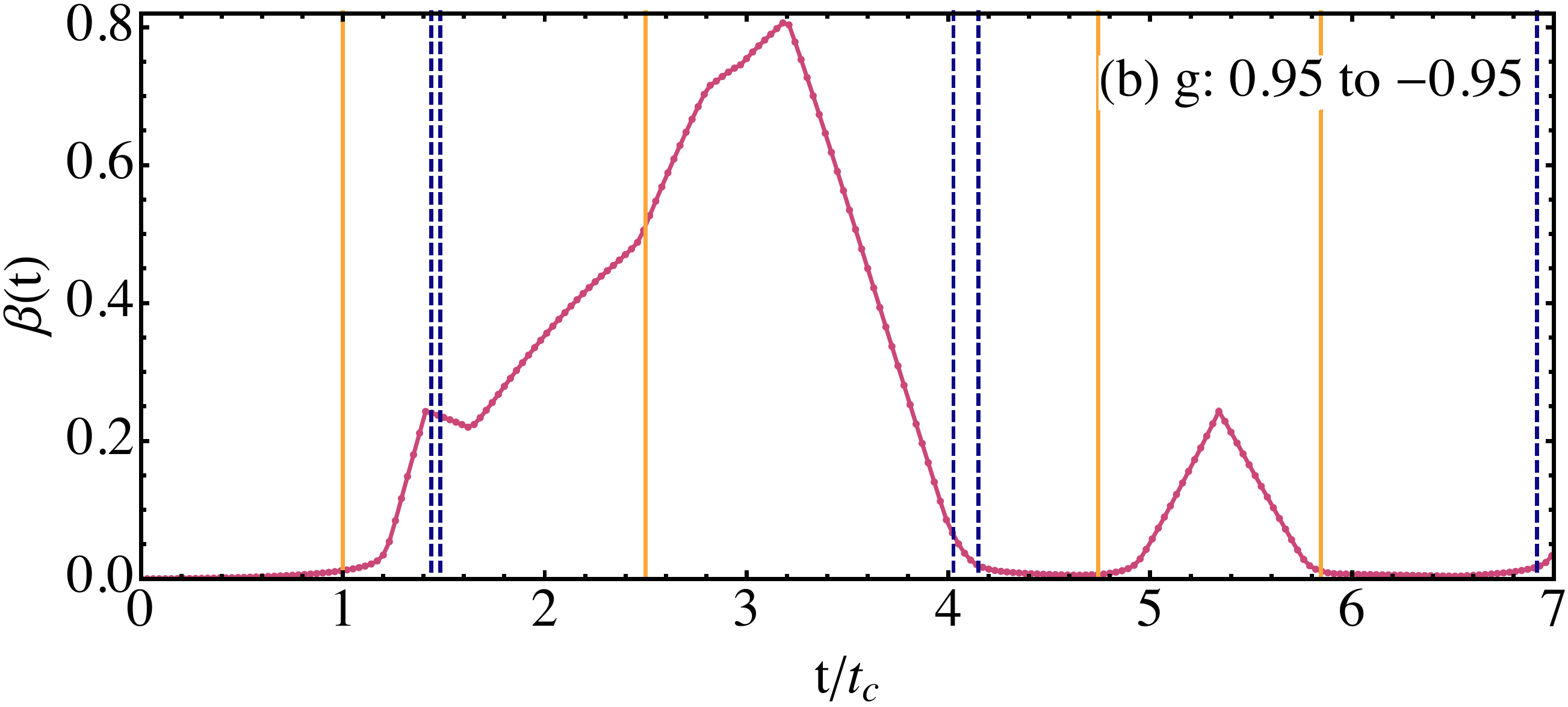}
\caption{(a) The smallest eigenvalues of $\mathbf{M}$, $\lambda_i(t)$, and (b) $\beta(t)$, for $N=4$, $N_w=160$, and $g_1=-g_2=0.95$ (trivial to non-trivial quench). For open boundary conditions. The orange lines show the location of the dynamical quantum phase transitions calculated for $\cos[\delta\phi_k]=0$ and (dashed) purple for $\cos[\delta\phi_k]\neq0$.}
\label{figure_10}
\end{figure}

\begin{figure}
\includegraphics*[width=0.99\columnwidth]{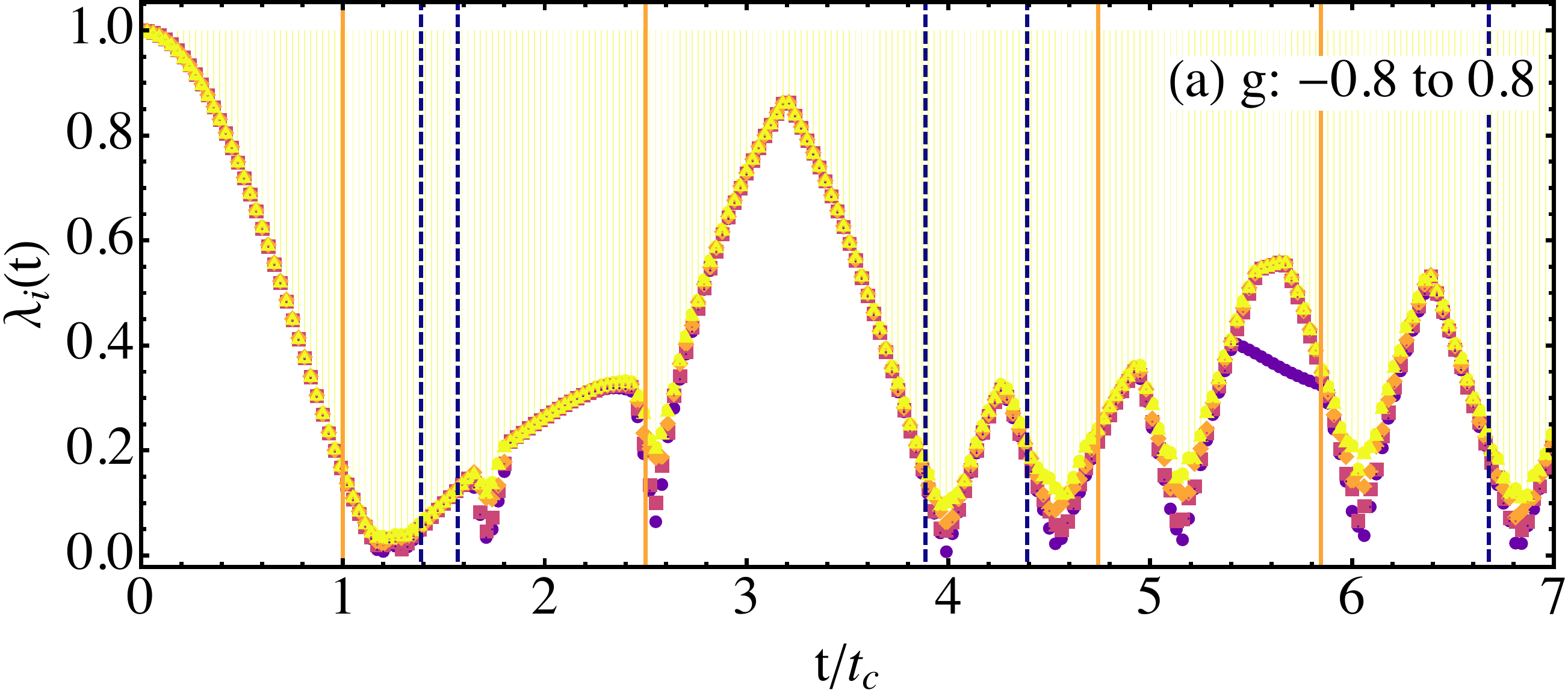}\\
\includegraphics*[width=0.99\columnwidth]{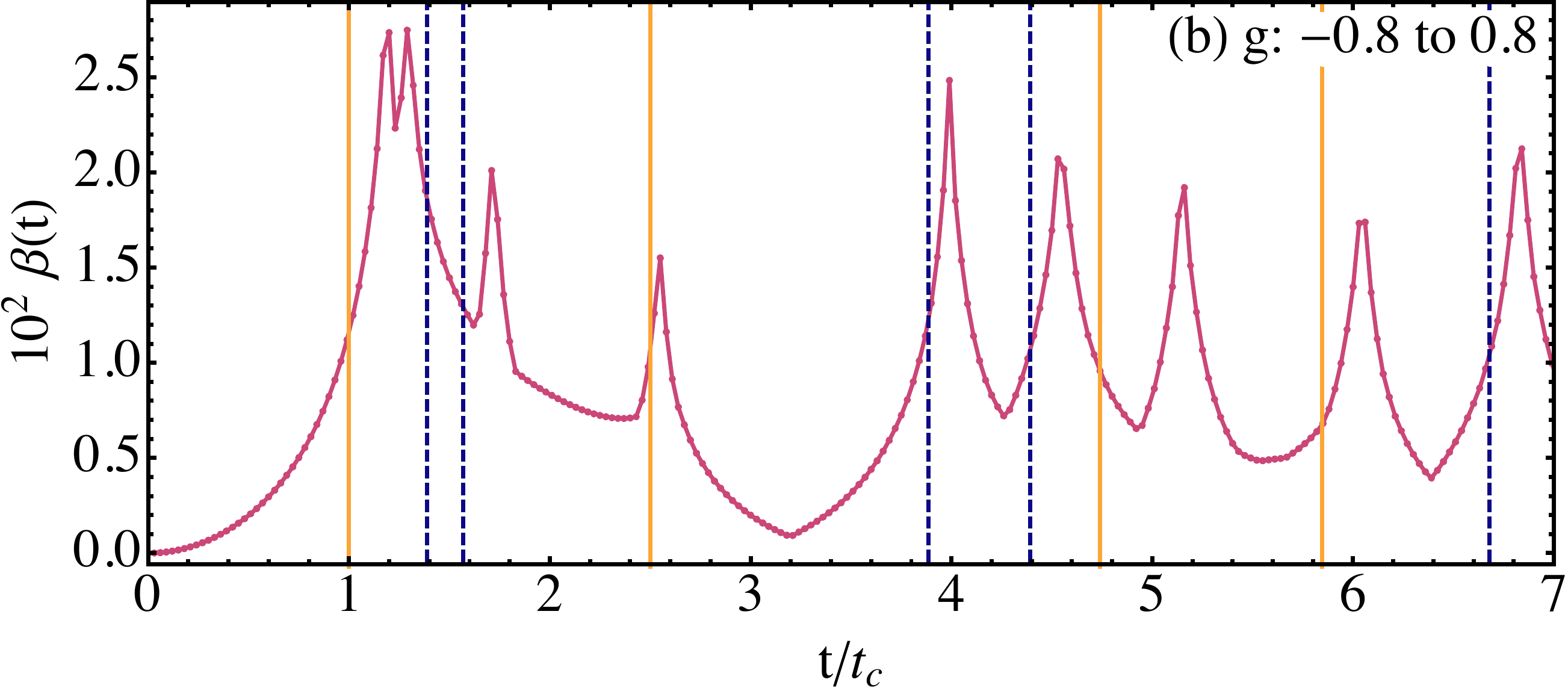}
\caption{(a) The smallest eigenvalues of $\mathbf{M}$, $\lambda_i(t)$, and (b) $\beta(t)$, for $N=4$, $N_w=160$, and $g_1=-g_2=-0.8$ (non-trivial to trivial quench). Note the rescaled $y$-axis. For open boundary conditions. The orange lines show the location of the dynamical quantum phase transitions calculated for $\cos[\delta\phi_k]=0$ and (dashed) purple for $\cos[\delta\phi_k]\neq0$.}
\label{figure_11}
\end{figure}

\begin{figure}
\includegraphics*[width=0.99\columnwidth]{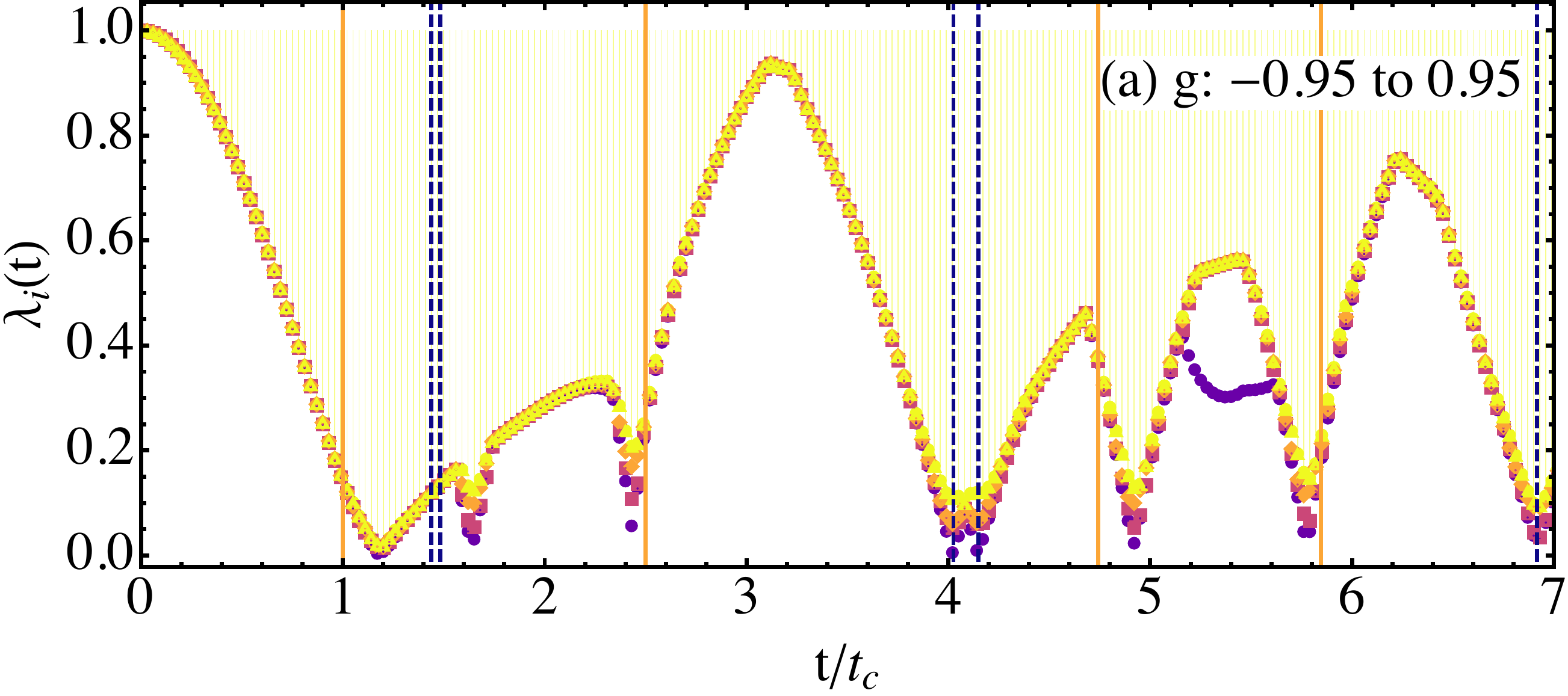}\\
\includegraphics*[width=0.99\columnwidth]{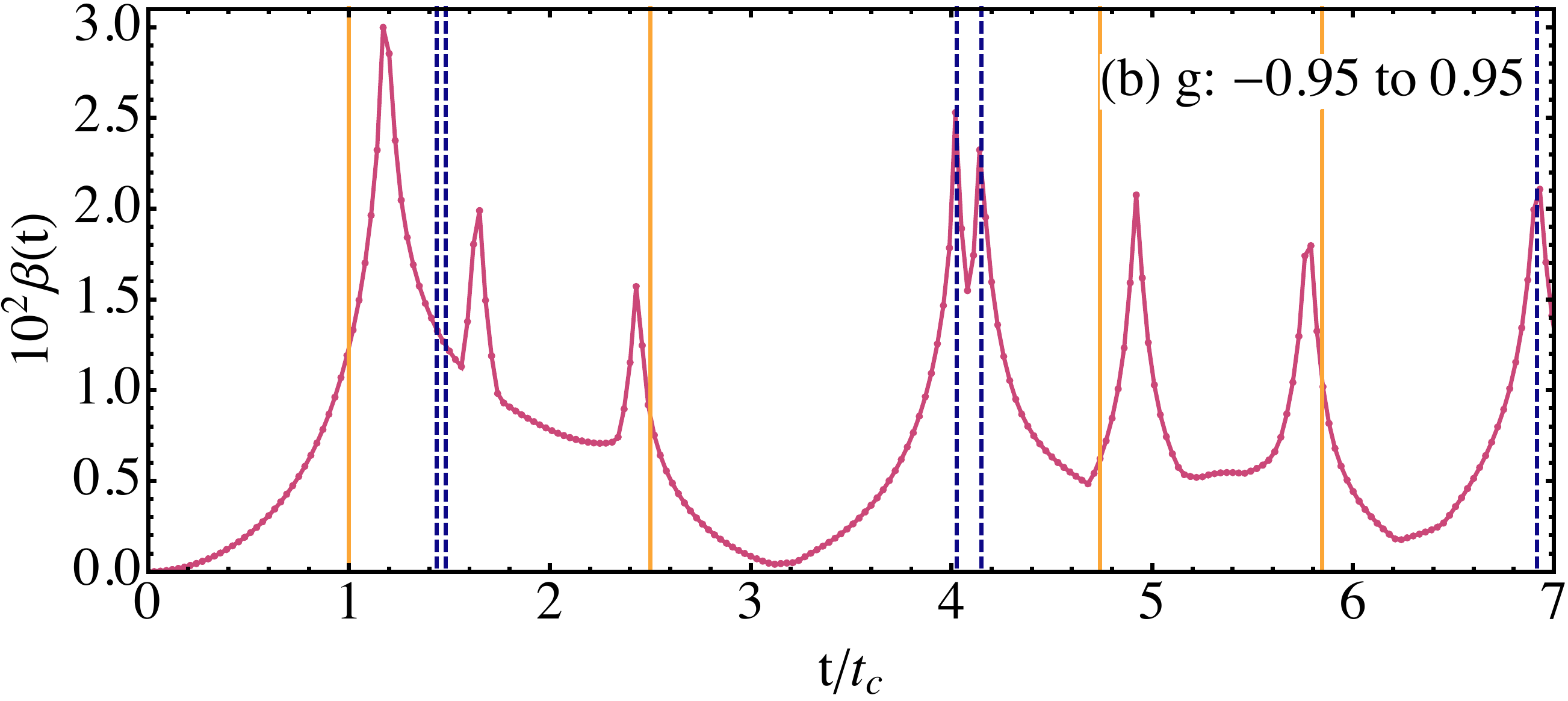}
\caption{(a) The smallest eigenvalues of $\mathbf{M}$, $\lambda_i(t)$, and (b) $\beta(t)$, for $N=4$, $N_w=160$, and $g_1=-g_2=-0.95$ (non-trivial to trivial quench). Note the rescaled $y$-axis. For open boundary conditions. The orange lines show the location of the dynamical quantum phase transitions calculated for $\cos[\delta\phi_k]=0$ and (dashed) purple for $\cos[\delta\phi_k]\neq0$.}
\label{figure_12}
\end{figure}

 To find the role played by the topologically protected edge states it is therefore necessary to perform a finite size scaling analysis to extract $l_B(t)$ from Eq.~\eqref{scaling}. Unfortunately for the generalized SSH model used here it would be extremely time consuming to numerically solve system sizes where the scaling sets in, and hence it is not possible to recover $l_B(t)$ by a finite size scaling analysis. Instead we rely on the behaviour of the lowest eigenvalues $\lambda_i(t)$ and their contribution to the finite size return rate $l(t)$. The return rate for a chain of length $N_w$ can be written as
\begin{equation}
	l(t) =\frac{1}{NN_w}\sum_{j=1}^{NN_w}\ln\left|\lambda_j(t)\right|\,.
\end{equation}
We can define the contribution to this from the lowest eigenvalues as
\begin{equation}
	\beta(t) =\frac{1}{NN_w}\sum_{j=1}^{2N}\ln\left|\lambda_j(t)\right|\,.
\end{equation}
For the SSH chain it has been shown that $l(t)-l_0(t)\approx\beta(t)$. In Figs.~\ref{figure_09}, \ref{figure_10}, \ref{figure_12}, and \ref{figure_11}, we plot both the smallest eigenvalues $\lambda_i(t)$ and $\beta(t)$ for a series of quenches for $N=4$. In all cases the dynamical bulk-boundary correspondence holds, with eigenvalues pinned to zero between critical times only for quenches into the topologically non-trivial phase. In fact due to finite size effects the eigenvalues are only approximately zero. We note that there appears to be no clear pattern for which critical time an eigenvalue goes to zero and for which it leaves zero. Even the different origins of the critical times does not play an organising role. I.e.~an eigenvalue can be pinned to zero at an orange line, $\cos[\delta\phi_k]=0$, and leave zero at a purple dashed line, $\cos[\delta\phi_k]\neq0$, see for example Fig.~\ref{figure_10}(a).

\begin{figure}
\includegraphics*[width=0.99\columnwidth]{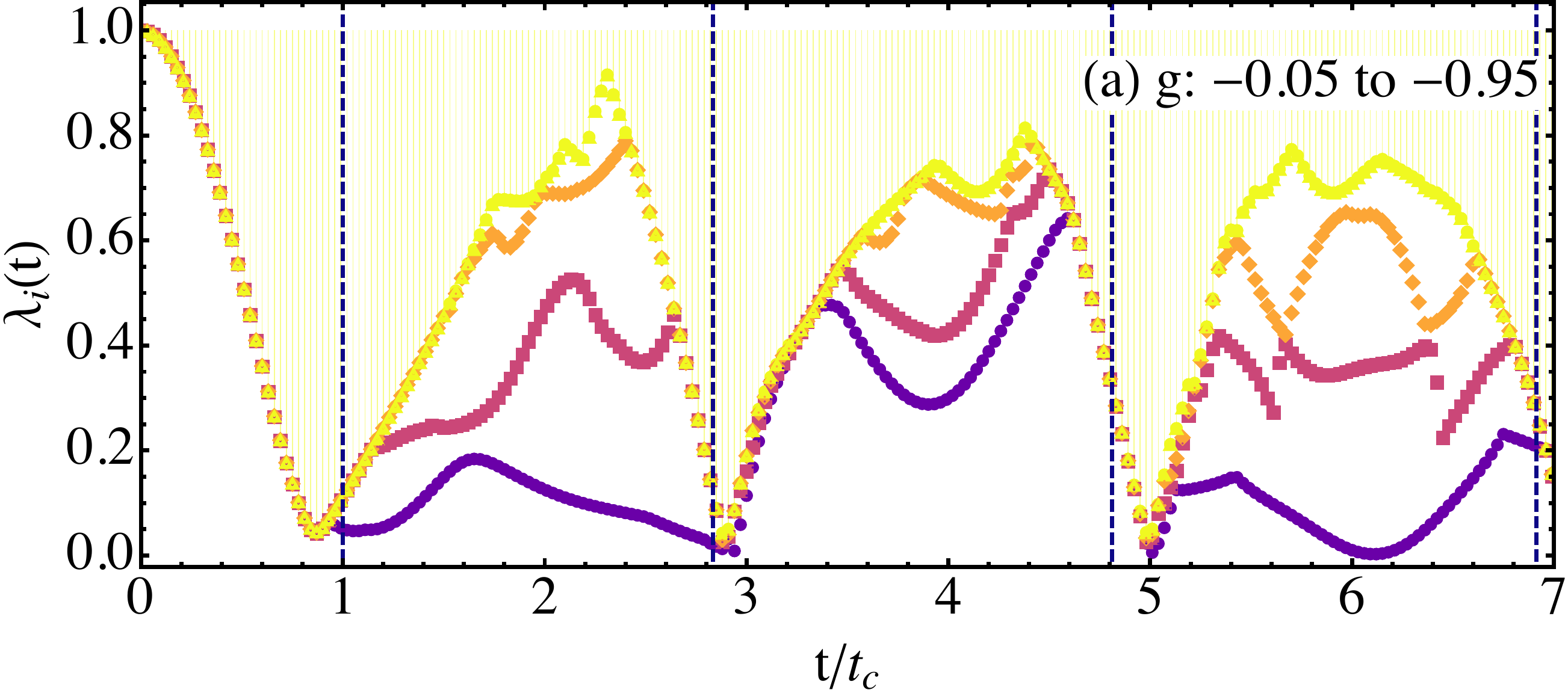}\\
\includegraphics*[width=0.99\columnwidth]{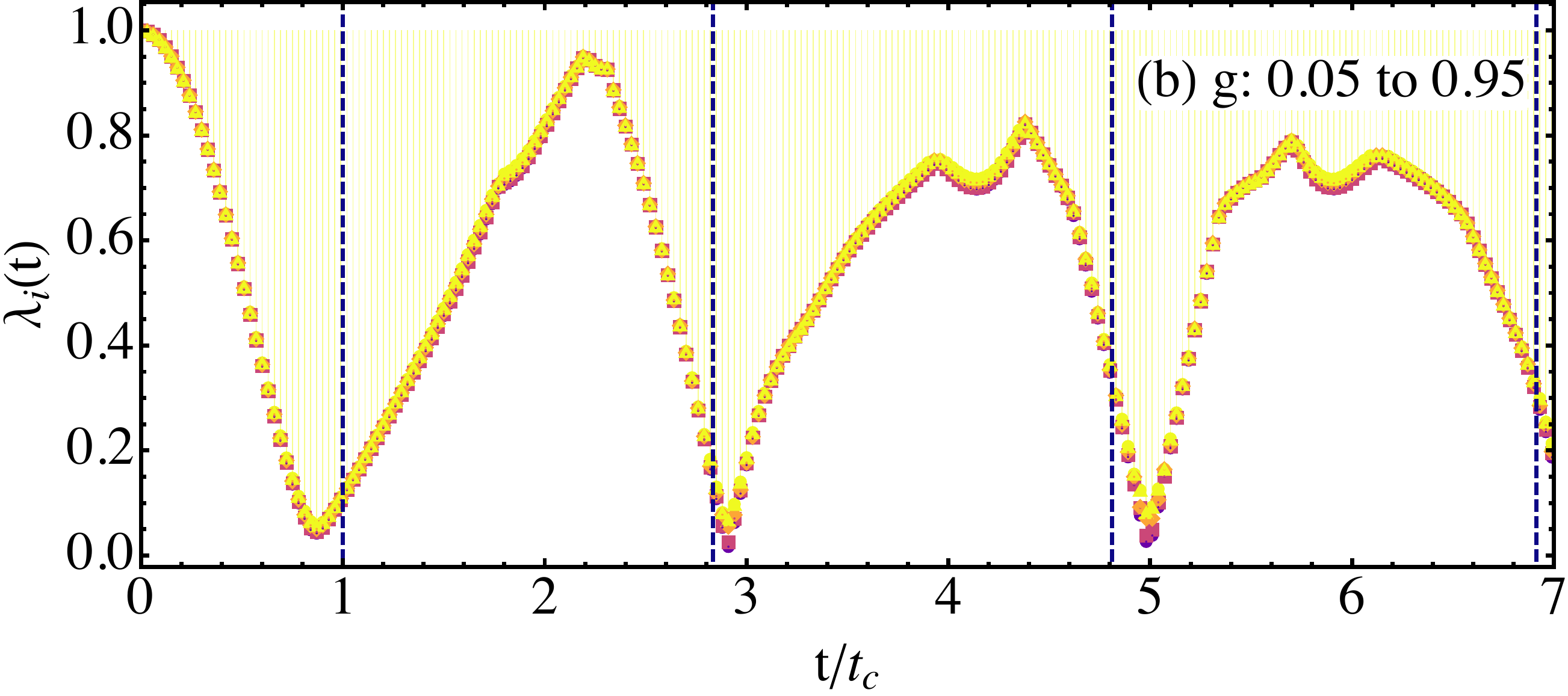}
\caption{The smallest eigenvalues of $\mathbf{M}$, $\lambda_i(t)$, for $N=4$, $N_w=160$, $|g_1|=0.05$, $|g_2|=0.95$. Panel (a) $g_{1,2}<0$ (quench within the non-trivial phase), and panel (b) $g_{1,2}>0$ (quench within the trivial phase). For open boundary conditions. The (dashed) purple lines show the location of the dynamical quantum phase transitions calculated for $\cos[\delta\phi_k]\neq0$.}
\label{figure_13}
\end{figure}

Curiously this dynamical bulk-boundary correspondence still holds for quenches within a phase. In Fig.~\ref{figure_13} $\lambda_i(t)$ for a quench within the trivial phase, and within the non-trivial phase, are shown. We find eigenvalues pinned to zero only for the quench within the topologically non-trivial phase. This suggests that they are related to the role of the topologically protected edge states within the dynamics, but a specific mechanism explaining them remains elusive.

\section{Relation to the dynamical entanglement entropy}
\label{sec_ent}

Previously, a still not fully understood connection between the timescales of dynamical quantum phase transitions and dynamic entanglement entropy has been reported.\cite{Sedlmayr2018} Oscillations in the dynamical entanglement entropy following a quench had a period exactly twice the critical timescale of the dynamical quantum phase transitions. The entanglement entropy is defined as a von-Neumann entropy for a subsystem after tracing out the rest of the system:
\begin{equation}
 S_{\textrm{ent}}(t) =-\Tr \{\rho_A(t) \ln \rho_A(t)\}\,
\end{equation}
where $\rho_A(t)=\Tr_B |\Psi(t)\rangle\langle\Psi(t)|$ and $|\Psi(t)\rangle = \text{e}^{-\im H_2t}|\Psi_1\rangle$ is the time-evolved state. $A$ and $B$ are the two subsystems of the full system, and we make a partition exactly in the middle of the chain. As we have a Gaussian model the entanglement can be calculated from eigenvalues of the time dependent correlation matrix $\mathbf{\C}(t)=$ the entries of which are given by $\C_{ij}(t)=\langle\psi_1|\e^{\im H_2t}\hat{c}^\dagger_i\hat{c}_j\e^{-\im H_2t}|\psi_1\rangle$.\cite{Peschel2003} The entanglement entropy is then
\begin{equation}
\label{Sent}
S_{\textrm{ent}} = -\sum_j \left[ \eta_j\ln\eta_j +(1-\eta_j)\ln(1-\eta_j)\right] \, 
\end{equation}
with $\eta_j$ being the eigenvalues of $\mathbf{\C}(t)$ restricted to lattice sites within subsystem $A$.

\begin{figure}
\includegraphics*[width=0.99\columnwidth]{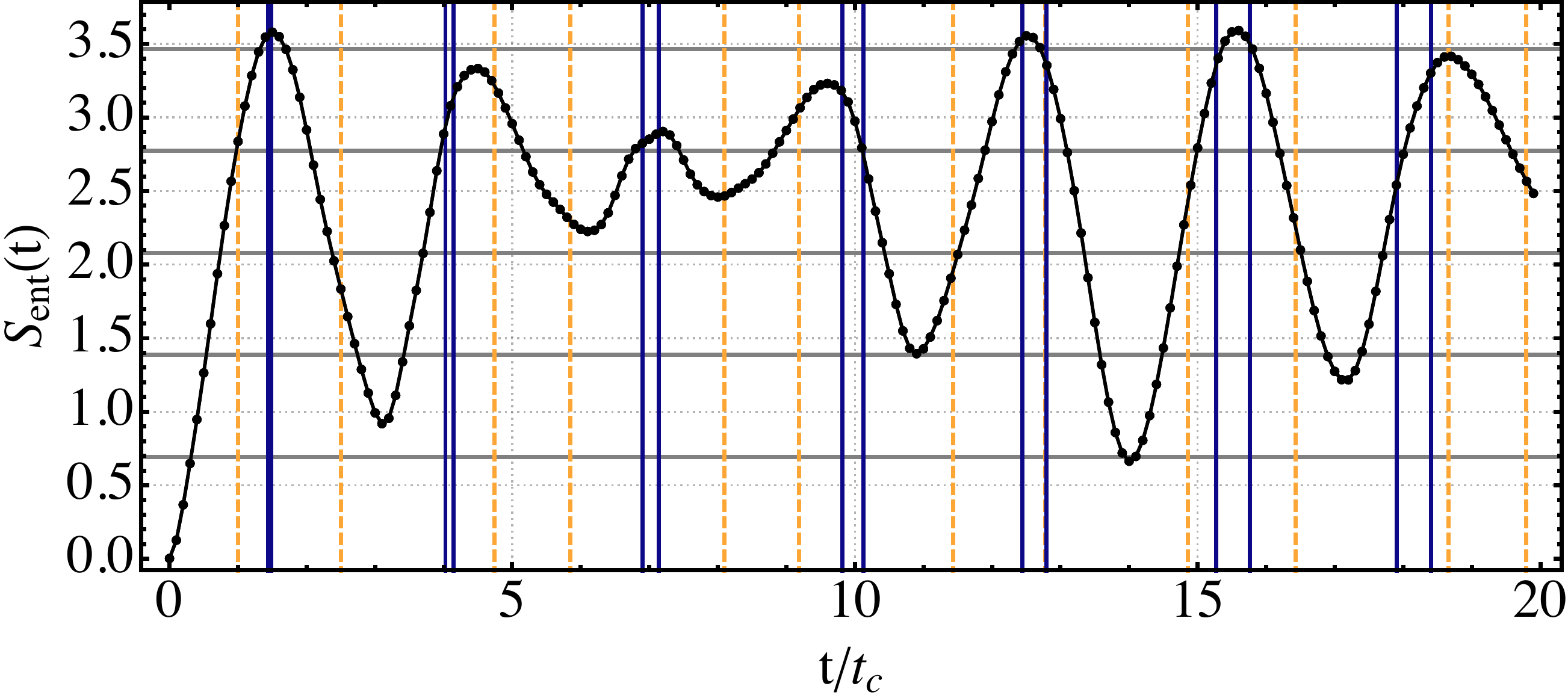}\\
\includegraphics*[width=0.99\columnwidth]{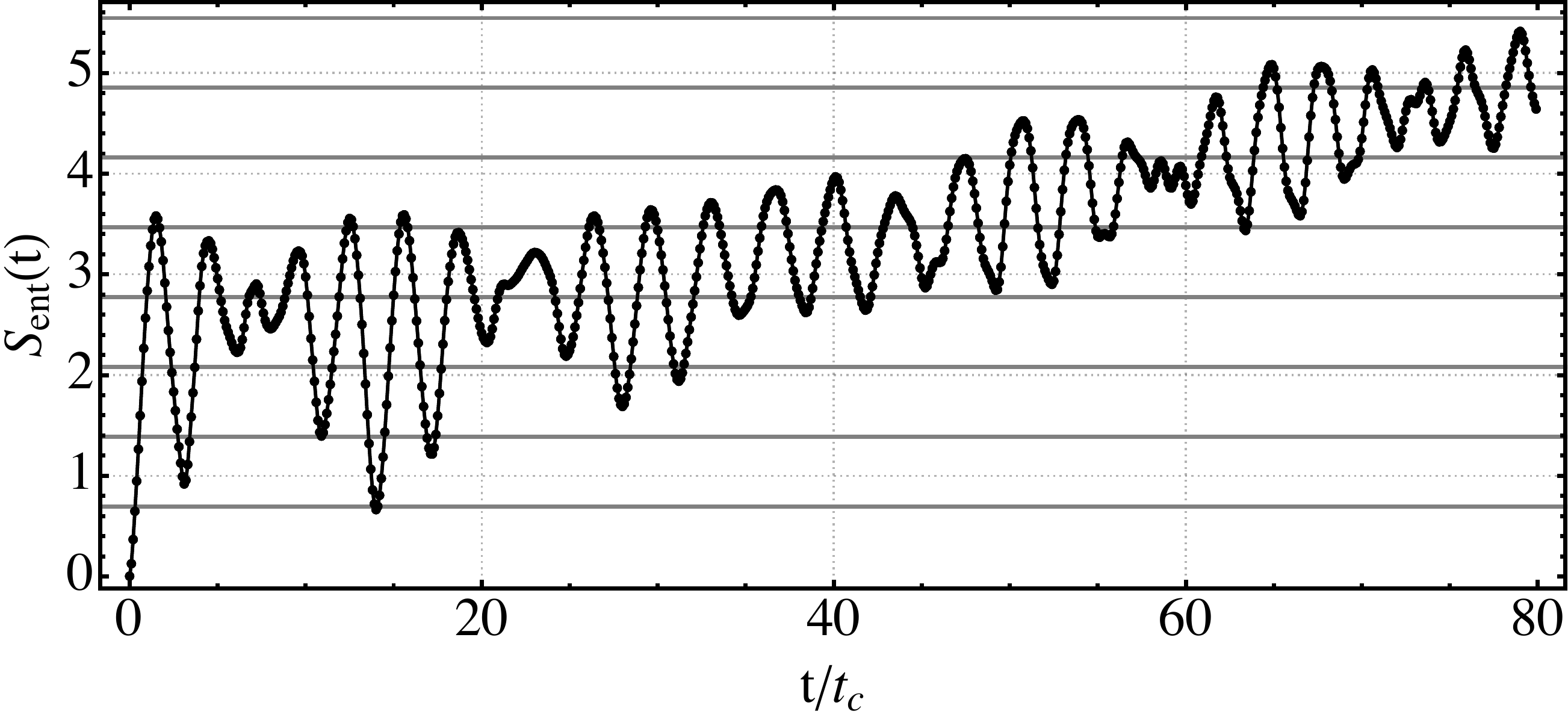}
\caption{The entanglement entropy for $N=4$ and $g_1=-g_2=0.95$. The orange lines show the location of the dynamical quantum phase transitions calculated for $\cos[\delta\phi_k]=0$ and (dashed) purple for $\cos[\delta\phi_k]\neq0$. Horizontal grey lines in (b) show $n\ln2$ for $n=1,2,\ldots,8$.}
\label{figure_14}
\end{figure}

\begin{figure}
\includegraphics*[width=0.99\columnwidth]{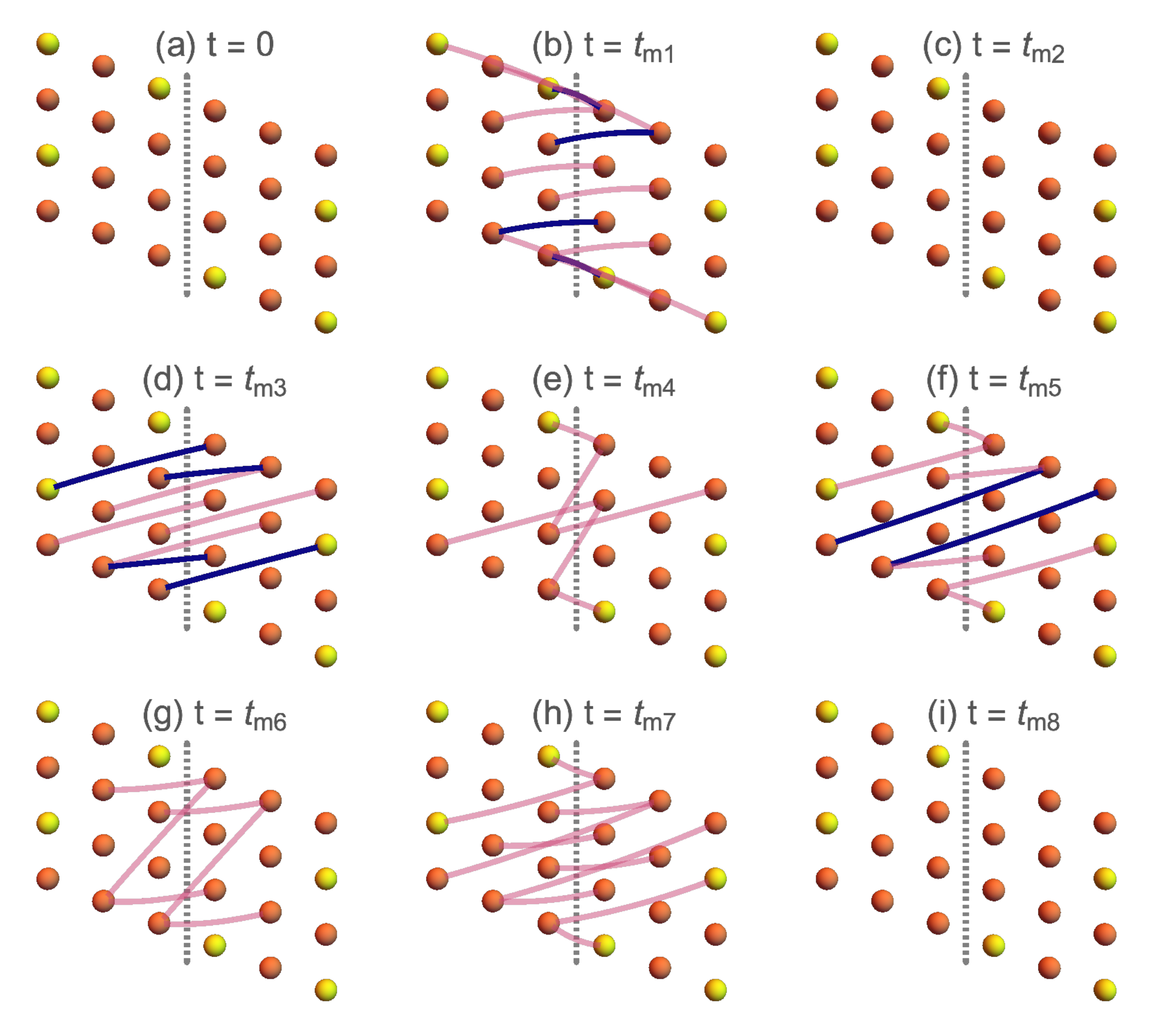}
\caption{The correlation matrix for $N=4$ and $g_1=-g_2=0.95$. If $C_{ij}>0.3$ then the connection is dark. If $0.2<C_{ij}<0.3$ then the connection is light. Only connections across the divide from the entanglement entropy are shown. No other connections are shown here. $t_{mi}$ are times of the maxima and minima of the entanglement entropy. A strong bond across the partition (the dashed grey line) contributes $\ln 2$ to the entanglement entropy. Yellow sites are the locations where the edge states are localized.}
\label{figure_15}
\end{figure}

In Fig.~\ref{figure_14} the entanglement entropy for a quench from $g_1=0.95$ to $g_2=-0.95$ is shown. No obvious connection between the critical times of the dynamical quantum phase transitions and the oscillations of the entanglement entropy is visible. However in this case the aperiodic nature of the critical times could obscure any clear connection. The timescales themselves remain similar. The entanglement entropy begin at zero, as in the state at $t=0$ there are no correlations between subsystems $A$ and $B$, see Fig.~\ref{figure_15}. The subsequent peaks and troughs of the entanglement entropy correspond to the development and reduction in correlations between sites in subsystems $A$ and $B$.

\begin{figure}
\includegraphics*[width=0.99\columnwidth]{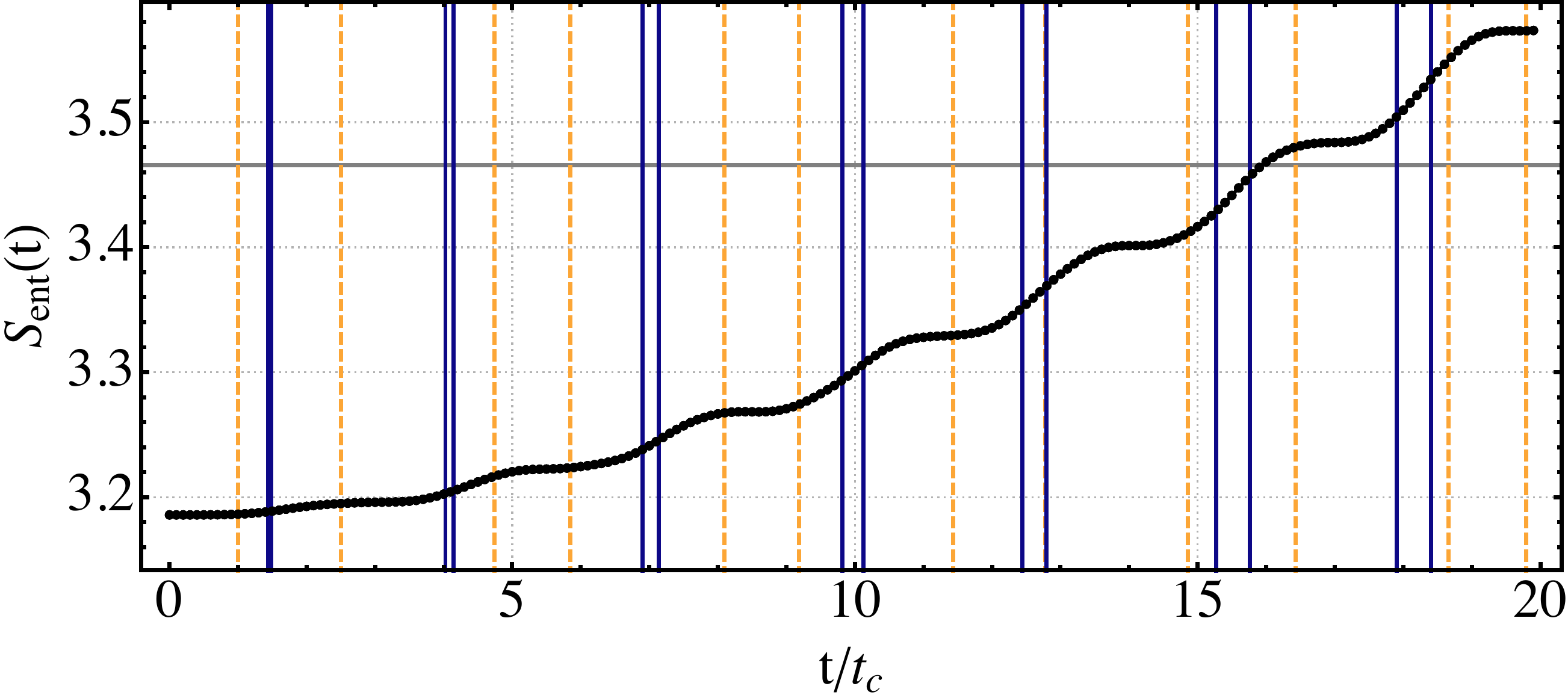}\\
\includegraphics*[width=0.99\columnwidth]{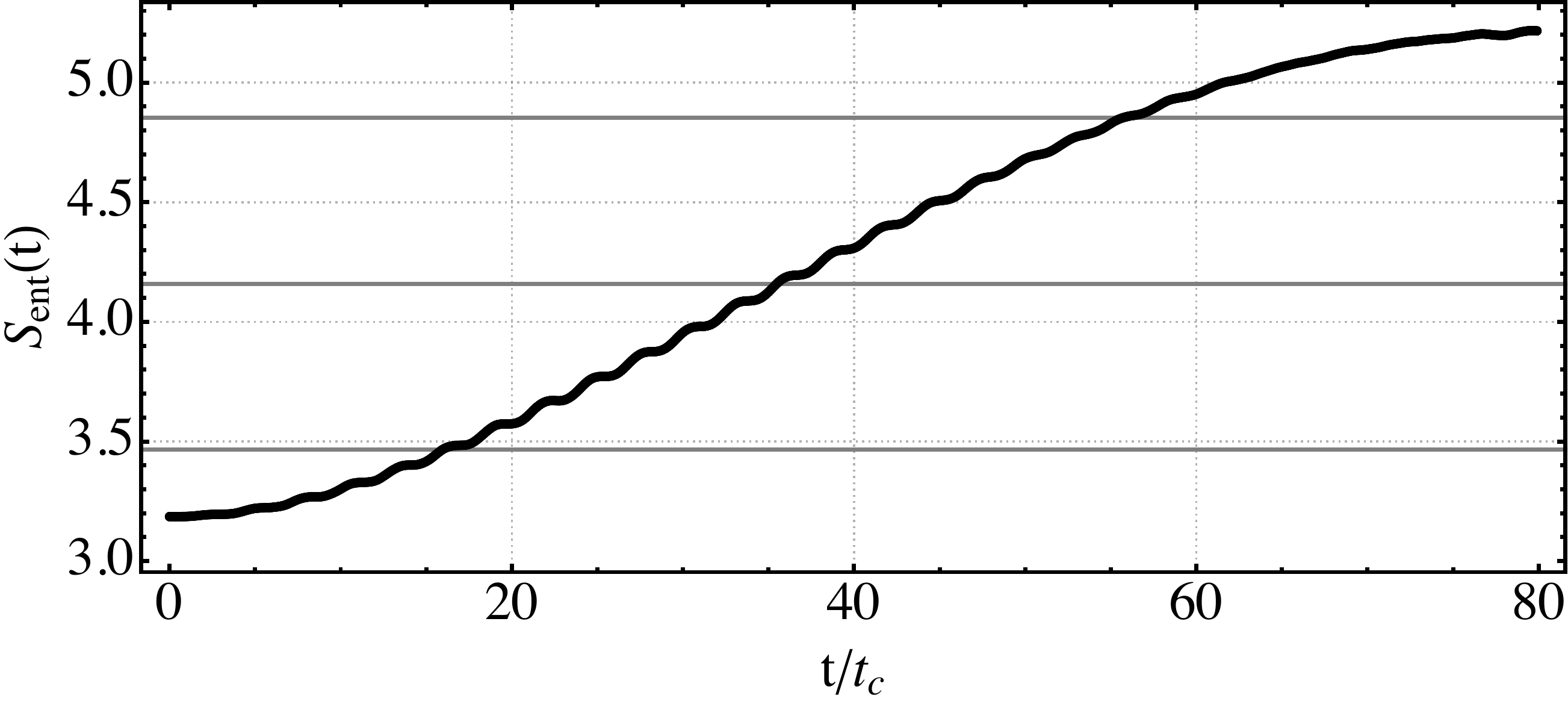}
\caption{The entanglement entropy for $N=4$ and $g_1=-g_2=-0.95$. The orange lines show the location of the dynamical quantum phase transitions calculated for $\cos[\delta\phi_k]=0$ and (dashed) purple for $\cos[\delta\phi_k]\neq0$. Horizontal grey lines in (b) show $n\ln2$ for $n=2,3,4$.}
\label{figure_16}
\end{figure}

\begin{figure}
\includegraphics*[width=0.99\columnwidth]{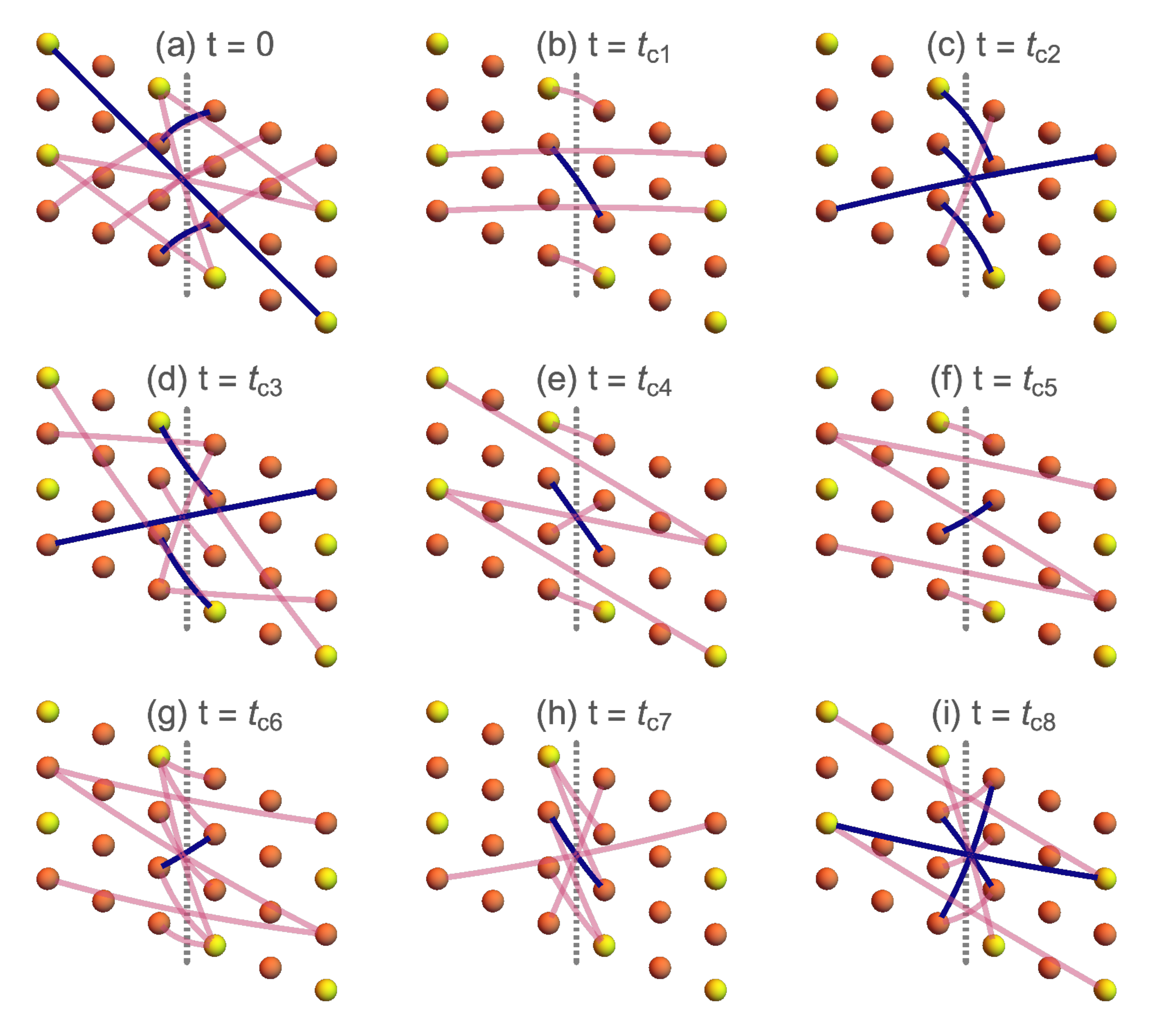}
\caption{The correlations for $N=4$ and $g_1=-g_2=-0.95$ at the critical times of the dynamical quantum phase transitions. If $C_{ij}>0.3$ then the connection is dark. If $0.2<C_{ij}<0.3$ then the connection is light. Only connections across the divide from the entanglement entropy are shown. A strong bond across the partition (the dashed grey line) contributes $\ln 2$ to the entanglement entropy. Yellow sites are the locations where the edge states are localized.}
\label{figure_17}
\end{figure}

For a quench which begins in the topologically non-trivial phase there is long range entanglement between the edge states,\cite{CamposVenuti2007a} which can be seen at $t=0$ in the entanglement entropy, see Fig.~\ref{figure_16}. Each strong bond cut contributes $\ln 2$ to the entanglement entropy, see Fig.~\ref{figure_17}. The subsequent dynamics of the entanglement entropy are caused by a reshuffling of the long range correlations, and a slow build up of the background entanglement. However, no clear connection with the critical times of the phase transition is evident

\section{Fidelity}
\label{sec_fidelity}

To further analyse the topological phase transition we consider the fidelity susceptibility across the transition. The fidelity between two states is defined as
\begin{equation}
	F(g,h)=|\langle\Psi(g)|\Psi(g+h)\rangle|\,,
\end{equation}
with $F(g,0)\equiv 1$. This measures the similarity between two ground states $H(g)|\Psi(g)\rangle=E(g)|\Psi(g)\rangle$. As before we consider half-filling and the Hamiltonian \eqref{ham}. The fidelity should drop across a (topological) phase transition where the ground states are now very dissimilar. However, similar to the Anderson orthogonality catastrophe\cite{Anderson1967} we expect a zero overlap in the thermodynamic limit and therefore consider
\begin{equation}
	\label{fidelity}
	f(g,h)=-\frac{1}{NN_w}\ln F(g,h)\,.
\end{equation}
We have additionally normalized by $N$, the number of atoms in a unit cell. Now as $f(g,0)\equiv 0$ is a minimum, we can expand the fidelity as $f(g,h)=\chi(\delta)h^2 +\mathcal{O}(h^3)$ with the fidelity susceptibility defined as 
\begin{equation}
	\label{fidelity_susci}
	\chi(g)=\frac{1}{2}\frac{\partial^2 f}{\partial h^2}\bigg|_{h=0}=-\frac{1}{NN_w}\lim_{h\to 0}\frac{\ln F(g,h)}{h^2}\,.
\end{equation}
The fidelity susceptibility has universal scaling behaviour at a quantum phase transition\cite{Zanardi2006,Venuti2007,Rams2011,Sirker2010,Konig2016} and can therefore be used to characterise such a transition.

In an expansion in $1/N_w$ we can write
\begin{equation}
	\chi=\chi_0 +\frac{\chi_B}{N_w} +\mathcal{O}(N_w^{-2})\,.
\end{equation}
The boundary susceptibility $\chi_B$ shows evidence of the edge states in the topologically non-trivial phase\cite{Sirker2014} and can be extracted via a finite size scaling analysis. The bulk contribution, $\chi_0$, can be calculated analytically for all $N$. In the following we calculate $\chi$ for $N=4,6,8$ and $N_W=20,40,\ldots, 260$ from which we extract the bulk and boundary fidelity susceptibilities.

In order to calculate the bulk fidelity susceptibility in the thermodynamic limit we first note that for a system of non-interacting electrons we can write
\begin{equation}
	\label{Fidelity_overlap}
	F(g,h)=|\det A_{kl}(g,h))|\,,
\end{equation}
with 
\begin{equation}
	\label{Overlap_matrix}
	A_{kl}(g,h)=\langle\psi_k(g)|\psi_l(g+h)\rangle
\end{equation}
being the $M\times M$ single particle overlap matrix and $M$ being the number of particles in the ground state. Using the eigenstates \eqref{eigenstate} we can calculate the overlap matrix as
\begin{eqnarray}
	A^{kk'}_{\lambda\lambda'}(g,h)&=&\langle\psi_{k\lambda}(g)|\psi_{k'\lambda'}(g+h)\rangle
	\\\nonumber
	&=&\delta_{kk'}\frac{2}{N+1}\sum_{j=1}^N\sin\frac{\pi j \lambda}{N+1}\sin\frac{\pi j \lambda'}{N+1}
	\\\nonumber&&
	\qquad\times\e^{\im[\phi_{k}(g)-\im\phi_{k}(g+h)]j}\,.
\end{eqnarray}
The $\lambda$ are confined to the occupied negative energy bands. Expanding in powers of $h$ we find for the non-zero diagonal in momentum elements:
\begin{eqnarray}
	A^{kk}_{\lambda\lambda'}(g,h)&=&\left[\delta_{\lambda\lambda'}-\im h\partial_g\phi_k(g)m_{\lambda\lambda'}\right]
	\\\nonumber&&
	-\frac{h^2}{2}\left[\im\partial^2_g\phi_k(g)m_{\lambda\lambda'}+[\partial_g\phi_k(g)]^2n_{\lambda\lambda'}\right]
	\\\nonumber&&
	+\Order(h^3)\,,
\end{eqnarray}
where
\begin{equation}
	\{m_{\lambda\lambda'},n_{\lambda\lambda'}\}=\frac{2}{N+1}\sum_{j=1}^N\{j,j^2\}\sin\frac{\pi j \lambda}{N+1}\sin\frac{\pi j \lambda'}{N+1}\,.
\end{equation}
Substituting this into \eqref{fidelity_susci} we find
\begin{equation}
	\chi(g)=
	\frac{1}{2NN_w}\Tr[\partial_g\phi_k]^2\left[\mathbf{n}-\mathbf{m}^2\right]\,.
\end{equation}
The trace is over momentum and the $\lambda$ space. This results in
\begin{eqnarray}
	\chi_0(g)&=&\frac{1}{2N_w}\sum_k\frac{\sin^2 k}{\left[1+g^2+(1-g^2)\cos k\right]^2}\\\nonumber&&\qquad
	\times\frac{1}{N}\sum_{\lambda=1}^\frac{N}{2}\left[\mathbf{n}-\mathbf{m}^2\right]_{\lambda\lambda}\,,
\end{eqnarray}
for the fidelity susceptibility.

Now taking the thermodynamic limit $N_w\to\infty$, the momentum sum becomes
\begin{eqnarray}
	\lim_{N_w\to\infty}\frac{1}{2N_w}\sum_k\frac{\sin^2 k}{\left[1+g^2+(1-g^2)\cos k\right]^2}\qquad\qquad\\\nonumber
	=\int_0^{\frac{\pi}{2}}\frac{\ud k}{4\pi}\frac{\sin^2 2k}{(\cos^2 k+g^2\sin^2 k)^2}=\frac{1}{4|g|(1+|g|)^2}\,.
\end{eqnarray}
Then defining
\begin{equation}
	\Gamma_N\equiv\underbrace{\frac{1}{4N}\sum_{\lambda=1}^\frac{N}{2}n_{\lambda\lambda}}_{=\frac{(N+1)(2N+1)}{48}}-\frac{1}{4N}\sum_{\lambda,\lambda'=1}^\frac{N}{2}m_{\lambda\lambda'}m_{\lambda'\lambda}\,,
\end{equation}
finally we have
\begin{equation}\label{bulksusc}
	\chi_0(g)=\frac{\Gamma_N}{|g|(1+|g|)^2}\,.
\end{equation}
For $N=2$ we recover the known result.\cite{Sirker2014}
We note here the first two values of interest of the prefactor $\Gamma_N$: $\Gamma_2=\frac{1}{32}$ and $\Gamma_4=\frac{9}{160}$. More are given in Appendix \ref{app_exp}.

In Fig.~\ref{figure_18} the bulk fidelity susceptibility extracted form scaling is compared to the exact result. Eq.~\eqref{bulksusc} suggests that the bulk fidelity susceptibilities can all be collapsed onto a single curve by an appropriate scaling. In Fig.~\ref{figure_18} we plot $\chi_0(g)\Gamma_2/\Gamma_N$, which collapses the susceptibilites onto the result for the SSH model with good agreement.

\begin{figure}
\includegraphics*[width=0.99\linewidth]{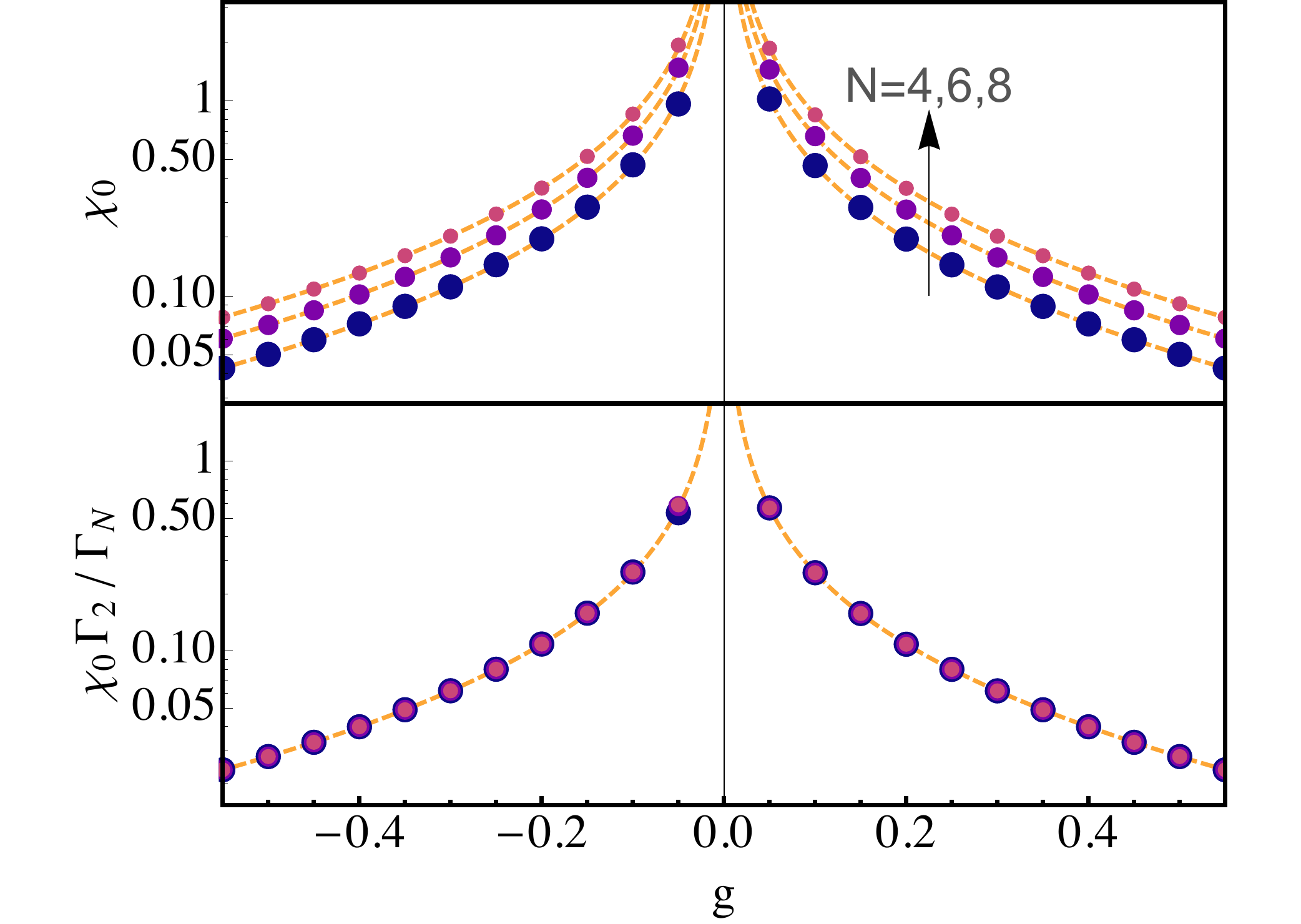}
\caption{(a) The bulk fidelity susceptibility $\chi_0(g)$ extracted from numerics (points) compared to the analytical result (dashed lines).(b) By rescaling $\chi_0(g)\Gamma_2/\Gamma_N$ all bulk results are collapsed onto the result for the SSH model.}
\label{figure_18}
\end{figure}

\begin{figure}
\includegraphics*[width=0.99\linewidth]{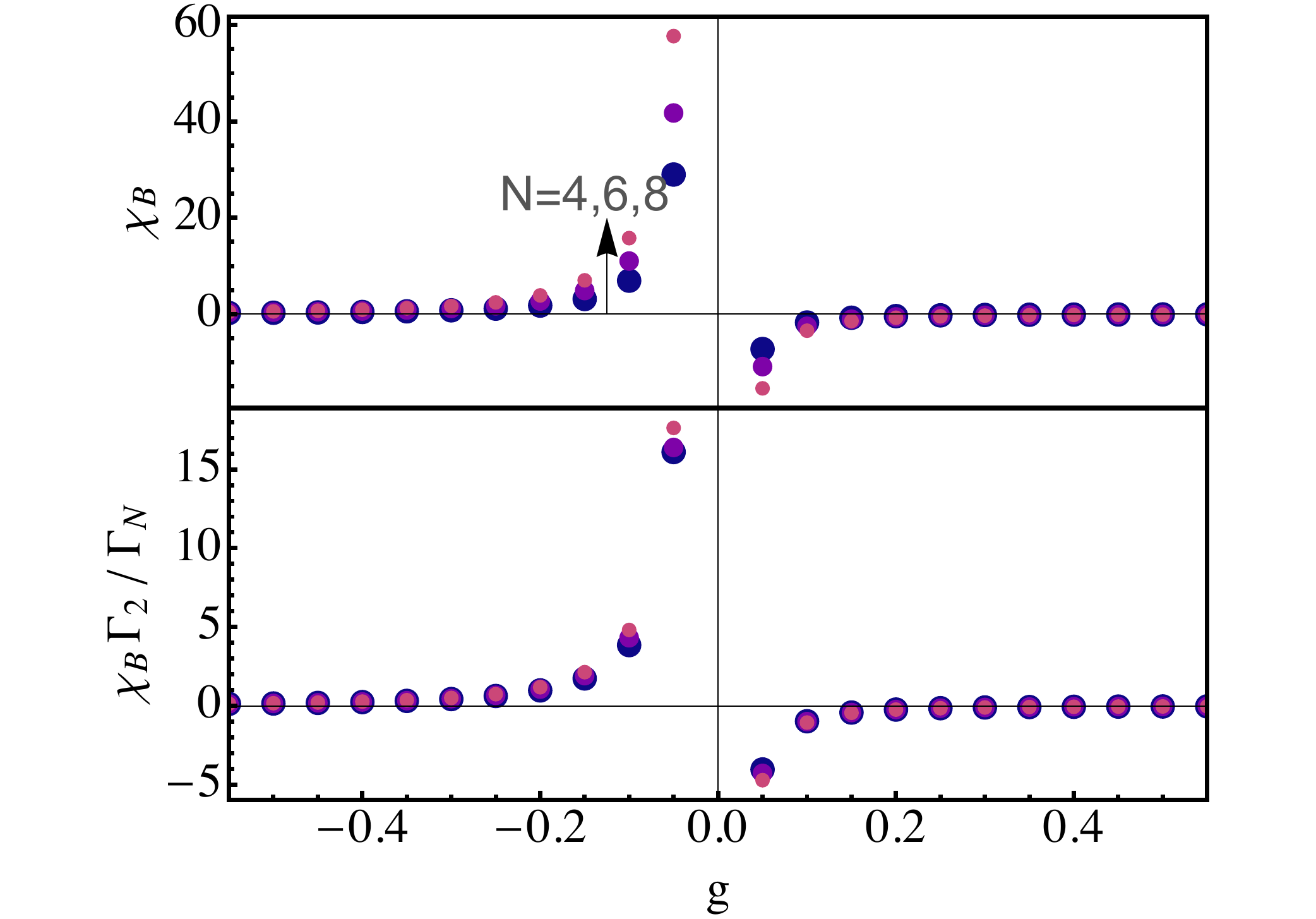}
\caption{(a) The boundary fidelity susceptibility $\chi_B(g)$ extracted from numerics (points) compared to the analytical result (dashed lines).(b) By rescaling $\chi_B(g)\Gamma_2/\Gamma_N$ all boundary results are collapsed onto the result for the SSH model to a reasonable agreement.}
\label{figure_19}
\end{figure}

Unexpectedly the exact same scaling works for the boundary contributions as well. In Fig.~\ref{figure_19} we first plot the results extracted from the finite size scaling. In all cases the strong asymmetry between the boundary susceptibility for $g<0$ and $g>0$ is caused by the presence of the edge states. By rescaling in the same way as for the bulk terms, $\chi_B(g)\Gamma_2/\Gamma_N$ the boundary contributions also fit to a single curve to good agreement. Deviations from this for small $|g|$ are possibly caused by limitations to the finite size scaling, but we can not rule out that these terms really deviate in that regime. The contribution from the edge states thus seems to follow the same pattern as the bulk susceptibility as a function of $N$. A comparison of the fidelity susceptibility for finite $N$ suggests that this rescaling hold at all orders in $N^{-1}$.

\section{Conclusions}
\label{sec_con}

In conclusion, we have introduced and thoroughly investigated the dynamical properties following a quench of a generalized SSH model. This model possesses a simple topological phase diagram and can be analytically solved, but nonetheless is a multiband topological insulator with rich dynamical features. We find non-periodic dynamical quantum phase transitions following a quench which arise from two distinct conditions, only one of which relates to the topology of the initial ground state and time evolving Hamiltonian. This first one is the equivalent of that previously discovered for two-band topological insulators, and here leads to quasiperiodic dynamical quantum phase transitions\cite{Karrasch2013}. The second, which also allows for dynamical quantum phase transitions when the quench is within a single topological phase, leads to aperiodic dynamical quantum phase transitions. This supplies further evidence that dynamical quantum phase transitions for quenches which do not cross equilibrium phase transitions are nonetheless ubiquitous once one moves away from minimal toy models.

Boundary contributions to the return rate confirm the previously reported dynamical bulk-boundary correspondence.\cite{Sedlmayr2018} For quenches in which the time evolving Hamiltonian is topologically non-trivial, large plateaus between critical times are present caused by eigenvalues of the Loschmidt matrix becoming pinned to zero. This remains the case even for quenches within a single topological phase.

We further investigated the dynamical entanglement entropy. Fluctuations in the entanglement entropy can be directly related to the shifting correlations between sites. Despite a close connection existing between these fluctuations and the critical timescale of the dynamical quantum phase transitions for the SSH model, here we see no evidence that the entanglement entropy fluctuations are related to the same timescale for this generalized model. 

The topological phase transition can be seen in the fidelity susceptibility, and an asymmetry to the boundary susceptibility is caused by the edge states. Interestingly both the bulk and boundary susceptibilities, and indeed any finite sized susceptibility, can be collapsed onto single curves by the same rescaling scheme. This requires further analysis in more general multiband models.

For the future a comprehensive understanding of the zero eigenvalues of the Loschmidt matrix for quenches into the topologically non-trivial regime would be interesting to see, but this requires further analysis on appropriately constructed models.

%%%%%%%%%%%%%%%%%%%%%%%%%%%%%%%%%%%%%

\acknowledgments

This research was supported in part by the National Science Centre (NCN, Poland) under grant 2018/29/B/ST3/01892 (NS) and in part by PLGrid Infrastructure (TM).

%%%%%%%%%%%%%%%%%%%%%%%%%%%%%%%%%%%%%

\appendix

\section{Some useful expressions}\label{app_exp}

We note here some useful sums for the open Fourier transform. The standard normalization condition is
\begin{equation}
	\frac{2}{N+1}\sum_{j}\sin^2\left[\frac{\pi j\lambda}{N+1}\right]=1\,.
\end{equation}
We will also make use of
\begin{equation}
	\frac{2}{N+1}\sum_{j}j\sin^2\left[\frac{\pi j\lambda}{N+1}\right]=\frac{N+1}{2}\,,
\end{equation}
and
\begin{eqnarray}\label{musum}
	4\sum_{\mu:\epsilon_{k\mu}<0}
	\sin\left[\frac{\pi\mu i}{N+1}\right]\sin\left[\frac{\pi\mu j}{N+1}\right]\qquad\\\nonumber=\frac{\sin\left[\frac{(i-j)\pi}{2}\right]}{\sin\left[\frac{(i-j)\pi}{2(N+1)}\right]}-\frac{\sin\left[\frac{(i+j)\pi}{2}\right]}{\sin\left[\frac{(i+j)\pi}{2(N+1)}\right]}\,,
\end{eqnarray}
which appears in Eq.~\eqref{corrmatrix}.

The first few values of $\Gamma_N$ for the bulk fidelity susceptibility, Eq.~\eqref{bulksusc}, are:
\begin{eqnarray}
	\Gamma_2&=&\frac{1}{32}\,,\\\nonumber
	\Gamma_4&=&\frac{9}{160}\,,\\\nonumber
	\Gamma_6&=&\frac{988-1285\cos[\frac{\pi}{7}]-1189\sin[\frac{\pi}{14}]+997\sin[\frac{3\pi}{14}]}{2352}\,,\\\nonumber
	\Gamma_8&=&\frac{1482-8045\cos[\frac{\pi}{9}]+6861\cos[\frac{2\pi}{9}]+7789\sin[\frac{\pi}{18}]}{5184}\,.
\end{eqnarray}

\section{Eigenstate densities}\label{app_densities}

\begin{figure}
\includegraphics*[width=0.8\linewidth]{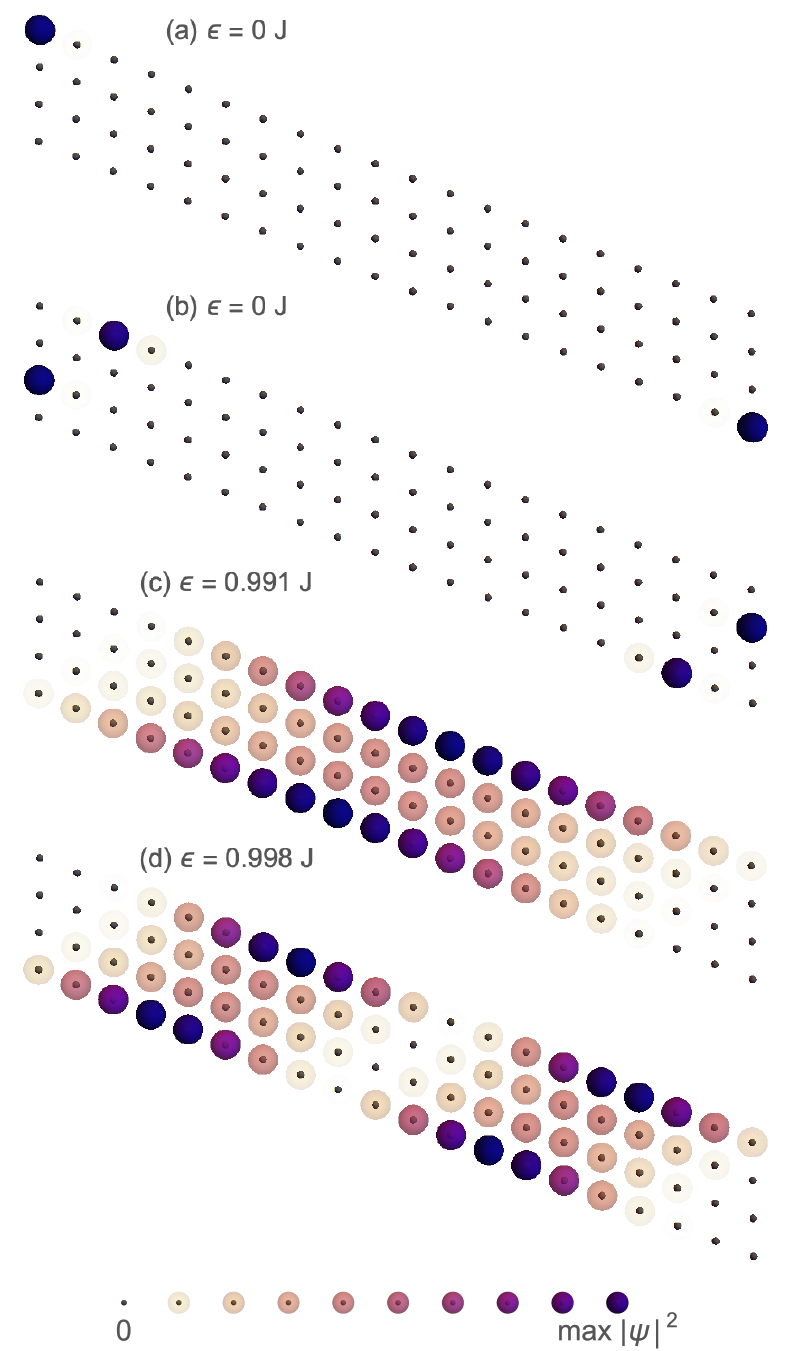}
\caption{Densities for eigenstates of the generalized SSH wire with $N=4$ and $g=-0.8$. The lowest 4 positive energy states are shown, including 2 positive ``zero energy'' edge states, (a) and (b), along with the 2 lowest energy bulk states (c) and (d).}
\label{figure_02_4}
\end{figure}

In Fig.~\ref{figure_02_4} we show the densities of the 4 smallest positive energy eigenstates of the chain for $N=4$ and $g=-0.8$. As for $N=6$ the edge states have a distinct pattern.

\section{Symmetries of the model}\label{app_symm}

The model given by Eq.~\eqref{ham} has a particle-hole symmetry, $\PH$ with $\PH^2=1$, and a `time-reversal' symmetry $\TR$, with $\TR^2=1$. Naturally it therefore also possesses their composite chiral symmetry $\PH\TR$. These satisfy the anti-commutation and commutation symmetries $[\PH,H]_+=0$ and $[\TR,H]_-=0$ respectively. Explicitly these are
\begin{equation}
	\left[\PH\right]_{ij}=(-1)^{j-1}\delta_{ij}\hat K
\end{equation}
for particle-hole symmetry, and
\begin{equation}
	\left[\TR\right]_{ij}=\hat K
\end{equation}
for `time-reversal' symmetry, where $\hat K$ is the complex conjugation operator. The chiral symmetry, $\mathcal{S}$, is therefore $[\mathcal{S},H]_+=0$ with
\begin{equation}
	\left[\mathcal{S}\right]_{ij}=\left[\PH\TR\right]_{ij}=(-1)^{j-1}\delta_{ij}\,.
\end{equation}
For $N=2$ these become the standard symmetry operators for the SSH model. Note that $\mathcal{S}$, $\PH$, and $\TR$ all commute with each other.

\end{document}